\documentclass[useAMS,usenatbib]{mn2e}
\usepackage{graphicx}
\usepackage{rotating}
\usepackage{longtable}
\usepackage{ulem}
\def \aj {AJ}
\def \mnras {MNRAS}
\def \pasp {PASP}
\def \apj {ApJ}
\def \apjs {ApJS}
\def \apjl {ApJL}
\def \aap {A\&A}
\def \nat {Nature}

\begin{document}

\title[Observations of the Type Ibn supernova OGLE-2012-SN-006]{Massive stars exploding in a He-rich circumstellar medium. V. Observations of the slow-evolving SN Ibn OGLE-2012-SN-006}

\author[Pastorello et al.]{A. Pastorello,$^1$\thanks{andrea.pastorello@oapd.inaf.it} {\L}. Wyrzykowski,$^{2,3}$ S. Valenti,$^{4,5}$ J. L. Prieto,$^{6,7}$ S. Koz{\l}owski,$^{2}$
\newauthor    A. Udalski,$^2$ N. Elias-Rosa,$^1$ A. Morales-Garoffolo,$^8$ J. P. Anderson,$^{9}$ S. Benetti,$^1$
\newauthor     M. Bersten,$^{10,11,12}$ M. T. Botticella,$^{13}$ E. Cappellaro,$^1$ G. Fasano,$^1$ M. Fraser,$^{14}$ 
\newauthor    A. Gal-Yam,$^{15}$ M. Gillone,$^{16}$ M. L. Graham,$^{4,5}$ J. Greiner,$^{17}$ S. Hachinger,$^{18,1}$ 
\newauthor   D. A. Howell,$^{4,5}$ C. Inserra,$^{19}$  J. Parrent,$^{4,20}$ A. Rau,$^{17}$ S. Schulze,$^{7,21}$
\newauthor   S. J. Smartt,$^{19}$ K. W. Smith,$^{19}$  M. Turatto,$^1$ O. Yaron,$^{15}$ D. R. Young,$^{19}$
\newauthor   M. Kubiak,$^{2}$  M. K. Szyma{\'n}ski,$^{2}$   G. Pietrzy{\'n}ski,$^{2,22}$ I. Soszy{\'n}ski,$^{2}$
\newauthor   K. Ulaczyk,$^{2}$ R. Poleski,$^{2,23}$  P. Pietrukowicz,$^{2}$  J. Skowron,$^{2}$ and P. Mr{\'o}z.$^{2}$
\\
$^{1}$ INAF - Osservatorio Astronomico di Padova, Vicolo dell' Osservatorio 5,  35122 Padova, Italy\\
$^{2}$ Warsaw University Observatory, Al. Ujazdowskie 4, 00-478 Warszawa, Poland\\
$^{3}$ Institute of Astronomy, University of Cambridge, Madingley Road, Cambridge CB3 0HA, UK\\
$^{4}$ Las Cumbres Observatory Global Telescope Network, Inc. Santa Barbara, CA 93117, USA\\
$^{5}$ Department of Physics, University of California Santa Barbara, Santa Barbara, CA 93106-9530, USA\\
$^{6}$ N\'ucleo de Astronom\'ia de la Facultad de Ingenier\'ia, Universidad Diego Portales, Av. Ej\'ercito 441, Santiago, Chile\\
$^{7}$ Millennium Institute of Astrophysics, Santiago, Chile \\
$^{8}$ Institut de Ci\`encies de l'Espai (CSIC-IEEC), Campus UAB,  Torre C5, 2a planta, 08193 Barcelona, Spain\\
$^{9}$ European Southern Observatory, Alonso de Cordova 3107, Vitacura, Santiago, Chile\\
$^{10}$ Instituto de Astrof\'isica La Plata, IALP (CCT La Plata), CONICET-UNLP, Paseo del Bosque s/n, 1900 La Plata, Argentina\\
$^{11}$ Facultad de Ciencias Astron\'omicas y Geof\'isicas, Universidad Nacional de La Plata, Paseo del Bosque s/n, 1900 La Plata, Argentina\\
$^{12}$ Kavli Institute for the Physics and Mathematics of the Universe (WPI), The University of Tokyo, Kashiwa, Chiba 277-8583, Japan\\
$^{13}$ INAF – Osservatorio Astronomico di Capodimonte, Salita Moiariello, 16 80131 Napoli, Italy  \\
$^{14}$ Institute of Astronomy, University of Cambridge, Madingley Road, Cambridge CB3 0HA, UK \\
$^{15}$ Department of Particle Physics and Astrophysics, Faculty of Physics, The Weizmann Institute of Science, Rehovot 76100, Israel\\ 
$^{16}$ School of Physics and Astronomy, University of Birmingham, Edgbaston, Birmingham B15 2TT, UK \\ 
$^{17}$ Max-Planck-Institut f\"ur extraterrestrische Physik, Giessenbachstrasse 1, 85748 Garching, Germany \\
$^{18}$ Institut f\"ur Theoretische Physik und Astrophysik, Universit\"at W\"urzburg, Emil-Fischer-Str. 31, 97074 W\"urzburg, Germany\\ 
$^{19}$ Astrophysics Research Centre, School of Mathematics and Physics, Queen's University Belfast, Belfast BT7 1NN, UK\\
$^{20}$ 6127 Wilder Lab, Department of Physics $\&$ Astronomy, Dartmouth College, Hanover, NH 03755, USA \\
$^{21}$ Instituto de Astrof\'{\i}sica, Facultad de F\'{i}sica, Pontificia Universidad Cat\'{o}lica de Chile, 306, Santiago 22, Chile\\ 
%$^{19}$ Millennium Institute of Astrophysics, Vicu\~{n}a Mackenna 4860, 7820436 Macul, Santiago, Chile\\
$^{22}$ Departamento de F\'isica, Universidad de Concepci\'on, Casilla 160-C, Concepci\'on, Chile\\
$^{23}$ Department of Astronomy, Ohio State University, 140 West 18th Avenue, Columbus, OH 43210, USA\\}
\date{Accepted XXXX Month XX. Received XXXX Month XX; in original form XXXX Month XX}

%\pagerange{\pageref{firstpage}--\pageref{lastpage}} \pubyear{201X}

\maketitle

\label{firstpage}

\clearpage

\begin{abstract}
We present optical observations of the peculiar Type Ibn supernova (SN Ibn) OGLE-2012-SN-006, discovered and monitored by the
$OGLE-IV$ survey, and spectroscopically followed by $PESSTO$ at late phases. 
Stringent pre-discovery limits constrain the explosion epoch with fair precision to $JD$ = 2456203.8 $\pm$ 4.0.
The rise time to the $I$-band light curve maximum is about two weeks.
The object reaches the peak absolute magnitude $M_I = -19.65 \pm 0.19$ on $JD = 2456218.1 \pm 1.8$. 
After maximum, the light curve declines for about 25 days with a rate of 4 mag 100d$^{-1}$.  
The symmetric $I$-band peak resembles that of canonical Type Ib/c supernovae (SNe), 
whereas SNe Ibn usually exhibit asymmetric and narrower early-time light curves.
Since 25 days past maximum, the light curve flattens with a decline rate slower than that 
of the $^{56}$Co to $^{56}$Fe decay, although at very late phases it steepens to approach that rate.
However, other observables suggest that the match with the  $^{56}$Co decay rate is a mere coincidence, and the radioactive decay
is not the main mechanism powering the light curve of OGLE-2012-SN-006.
An early-time spectrum is dominated by a blue continuum, with only a
marginal evidence for the presence of He I lines marking this SN Type. This spectrum shows 
broad absorptions bluewards than 5000\AA, likely O II lines, which are similar to spectral features observed in super-luminous 
SNe at early epochs.
The object has been spectroscopically monitored by $PESSTO$ from 90 to 180 days after peak,
and these spectra show the typical features observed in a number of SN 2006jc-like events, including a 
blue spectral energy distribution and prominent and narrow ($v_{FWHM} \approx 1900$ km s$^{-1}$) He I emission lines. This suggests
that the  ejecta are interacting with He-rich circumstellar material. The detection of broad ($10^4$ km s$^{-1}$) 
O I and Ca II features likely produced in the SN ejecta (including the [O I] $\lambda\lambda$6300,6364 doublet in the latest spectra) 
lends support to the interpretation of OGLE-2012-SN-006 as a core-collapse event.
\end{abstract}


\section{Introduction} \label{intro}

Supernovae of Type Ibn (SNe Ibn) are considered a rare group of stripped-envelope core-collapse (CC) events which interact with
H-depleted circumstellar material (CSM). The spectra of SNe Ibn are characterized by relatively narrow  lines of He I in emission 
\citep[hence the designation as ``Ibn'',][]{pasto08b}, with full-width at half maximum (FWHM) velocities ranging from several hundreds to a few 
thousands km s$^{-1}$. These features are thought to arise in the interactions between the SN ejecta and He-rich (and H-poor) CSM. 
However, weak H lines have been occasionally detected in the spectra of a few Type Ibn SNe, suggesting the 
presence of residual H in the CSM of -at least- a sub-sample of SNe Ibn \citep{pasto08a,smi12,pasto13a}.

The prototype of this family, SN 2006jc, has been associated with a precursor eruptive episode  which occurred a couple of years before
the SN explosion and reached a peak absolute magnitude $M_R$ $\approx$ -14 \citep{yam06,nak06,pasto07,fol07}.  
SN 2006jc, although discovered after the maximum light, was widely studied.
The brightest magnitude registered for this SN was $M_R$ = -18.3 \citep{pasto07}, and its optical light curve experienced a rapid post-peak decline 
(about 3.5 mag during the first 50 days after discovery). At later phases, the increasing decline rate of the optical light curve and the 
simultaneous near-infrared (NIR) excess were explained as the consequence of dust 
forming in a cool dense shell \citep[e.g.][]{smi08,seppo08}. 
\citet{tom08} and \citet{pasto08a} attempted to reproduce the evolution of SN 2006jc with radioactively powered models. They both favoured models
with a high kinetic energy (E$_k \sim$ 10$^{52}$ erg), and moderate masses of ejecta (M$_{ej}$ $\approx$ 4.5-5 M$_\odot$) and synthesized $^{56}$Ni (M$_{^{56}Ni} \approx$ 0.2-0.25 M$_\odot$).
Nonetheless, radioactively powered models with even lower E$_k$ and M$_{ej}$ provided decent matches with observations \citep[e.g., models A and B in][were obtained adopting the
following parameters: E$_k \sim$ 10$^{51}$ erg, M$_{ej} \leq$ 1 M$_\odot$ and M$_{^{56}Ni} \sim$ 0.25-0.40 M$_\odot$]{pasto08a}.
On the other hand, \citet{chu09} showed that radioactive models with modest M$_{ej}$/M$_{^{56}Ni}$ ratio were unrealistic for a Type Ib/c SN explosion,
and proposed that the observed parameters of SN 2006jc were best explained with strong
ejecta-CSM interaction occurring already at early phases. In fact, the interaction with a circumstellar shell with mass of a few hundredths M$_\odot$ 
%(with a pre-SN mass-loss rate of a few $\times$ 10${^-2}$ M$_\odot$ yr$^{-1}$)  
would fairly well explain the properties of SN 2006jc.
It is clear, however, that for an accurate modelling good data coverage is essential along the duration of the entire SN evolution, 
including the very early phases, not available for SN 2006jc.
The first opportunity to study a SN Ibn soon after its explosion was offered by SN 2010al, which initially 
showed an almost featureless spectrum with the shape of a hot blackbody \citep[the data of this object are presented in][]{pasto13a}. 

The proliferation of well-organized imaging surveys, with different depths and strategies, has significantly increased the number of transients discovered and, hence, the number
of Type Ibn candidates. The {\it Optical Gravitational Lensing Experiment} \citep[$OGLE$,][]{uda97} is a long-term  project carried out initially with the 1-m Swope telescope,
and later with the 1.3-m Warsaw University Telescope, both located at the Las Campanas Observatory (Chile). The fields monitored in the course of the different seasons 
of the search included the two Magellanic Clouds and the Magellanic Bridge.\footnote{A few additional fields were monitored by $OGLE$ in the Galactic bulge and disk.}
In the current fourth season \citep[$OGLE-IV$,][]{wyr14}, which started in 2009, the survey is making use of a 32 CCD mosaic camera covering a field of 1.4 square degrees and equipped with $V$ and 
$I$ filters. Originally conceived to study microlensing events \citep{pac86,pac91}, a by-product of the project is the discovery of
a huge number of transients and variable stars of all types, including SNe \citep{koz13}.

\begin{figure*}
\includegraphics[width=7.0in]{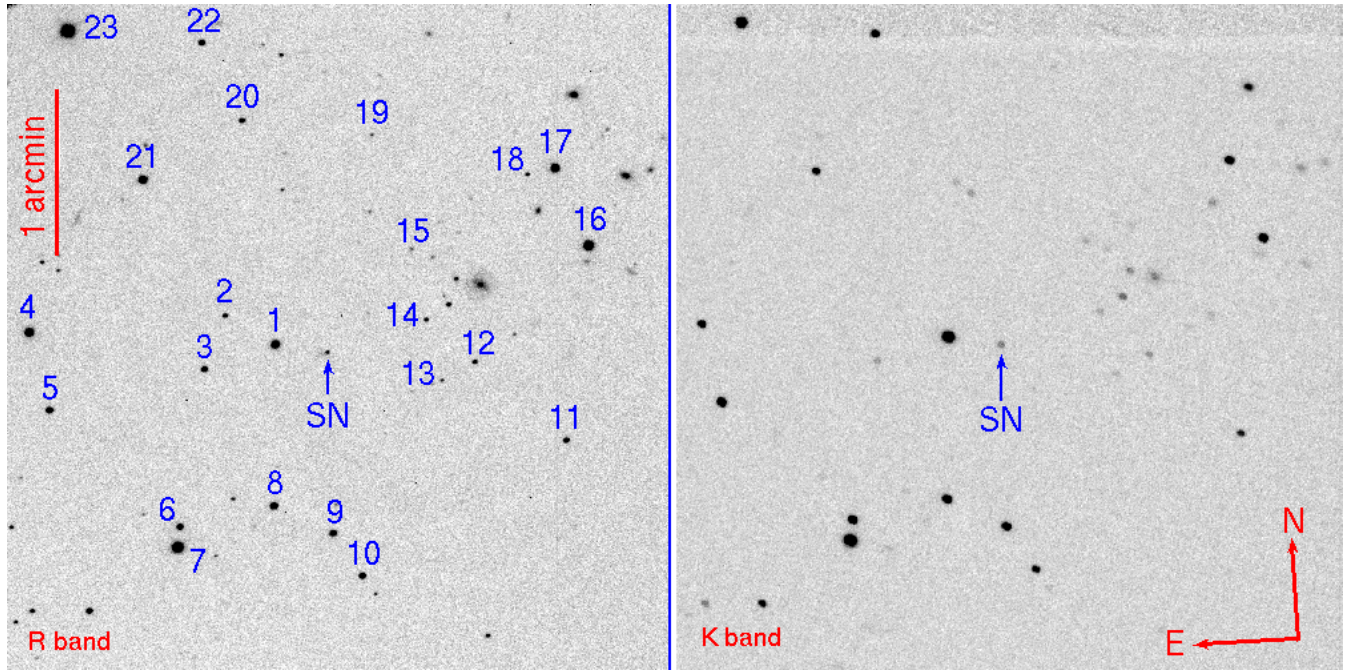}
\caption{Left: ESO-NTT EFOSC2 $R-$band image of OGLE-2012-SN-006 obtained on January 19, 2013; the numbers mark the reference stars used to calibrate the SN magnitude.
Right: ESO-NTT SOFI $Ks-$band image of OGLE-2012-SN-006 obtained on January 21, 2013.} 
\label{fig1}
\end{figure*}

The study of new types of stellar explosions is the main goal of the {\sl ``Public ESO Spectroscopic Survey of Transient Objects''} \citep[$PESSTO$,\footnote{{\it http://www.pessto.org}} see][]{sma13}, which is a 
public spectroscopic survey running at the 3.58-m New Technology Telescope (NTT) of the European Southern Observatory (ESO) in La Silla (Chile). The survey is planned
for four years (starting from April 2012) and is operative 90 nights per year.
The survey aims to classify a large number of transients and to monitor the most interesting objects with  EFOSC2 (optical) and SOFI (near-infrared).
Priority $PESSTO$ targets are the most unusual SNe, SN impostors or very nearby transients, for which detailed high-quality, multi-wavelength observational 
campaigns can be arranged.

A specific subproject has been defined within \textit{PESSTO} in order to study SNe Ibn \citep{pasto08b}.
Two Type Ibn SNe discovered by the $La Silla-QUEST$ variability survey and followed by $PESSTO$ (LSQ12btw and LSQ13ccw),
 are studied in a companion paper to the present work \citep{pasto13b}. 
A few months after LSQ12btw, a new SN Ibn candidate, OGLE-2012-SN-006,
 was discovered in the course of the $OGLE-IV$ survey \citep{wry12,wyr14}.
The object has coordinates  $RA = 3^h33^m34^s.79$ and $Dec = -74^o23'40''.1$ (J2000), and lies in a faint, anonymous host galaxy (Figure \ref{fig1}). 
According to \citet{wry12}, the object had a magnitude $I$ = 17.62 at maximum light. 
OGLE-2012-SN-006 was spectroscopically classified on 2013 January 10.2 UT by \citet{pri13} as a Type Ibn SN, similar to 
SN 2006jc. The host galaxy redshift, as estimated from the position of the narrow SN features, is z = 0.057$\pm$ 0.001. Adopting a Hubble constant value of 
H$_0$ = 73 $\pm$ 5 km s$^{-1}$ Mpc$^{-1}$ ($\Omega_{m}$ = 0.27, and $\Omega_{\Lambda}$ = 0.73), 
we obtain a luminosity distance of 244.5 $\pm$ 20.0 Mpc. 

\citet{pasto08b}, on the basis of a very small sample of objects, proposed that SNe Ibn might form a relatively homogeneous group of mildly 
interacting stripped-envelope SNe. However, recent discoveries suggest that SNe Ibn span a much wider range of properties than initially believed \citep{smi12,gor13,san13,pasto13a,pasto13b}. 
In this context, because of its slow evolving light curve (see Section \ref{lc}), OGLE-2012-SN-006 can be considered as an unprecedented addition to the Type Ibn SN zoo.

\section{Observations}  \label{obs}

Our photometric and spectroscopic data have been reduced using standard procedures in the IRAF environment.\footnote{IRAF is distributed by the National Optical Astronomy Observatory,
which is operated by the Association of Universities for Research in Astronomy (AURA) under cooperative agreement with the National Science Foundation.}

Optical imaging frames were first pre-reduced (i.e. overscan, bias and flat-field corrected, and trimmed in order to remove the unexposed regions of the image). 
In the near-infrared (NIR) photometry images we also removed the contribution of the bright NIR background from the science images. 
Sky images were  obtained by median-combining a number of dithered science frames, and were finally subtracted from each science image. 
In order to improve the signal to noise, the sky-subtracted science images were spatially registered  and finally combined.

Photometric measurements in the optical and NIR bands were obtained through the PSF-fitting technique. A template PSF was built  using stars
in the SN field. With this PSF model along with a low-order polynomial surface, we finally performed a fit to the SN and the underlying background.
$OGLE-IV$ photometry was obtained using the Difference Imaging Analysis (DIA), which is a template subtraction method adapted to the $OGLE$ data and detailed in \citet{wyr14} \citep[see also][]{woz00}.
The magnitudes of several stars in  the SN field (see Figure \ref{fig1}, left) were calibrated using observations of standard fields  of the \citet{lan92} catalogue. The magnitudes
obtained in photometric nights were used to estimate reliable average magnitudes for the entire local stellar sequence (23 stars), which was finally
used for the calibration of the SN photometry obtained during non-photometric nights. 

The NIR SN photometry was calibrated through a comparison with 14 stars of the optical sequence for which NIR magnitudes were available
in the The Two Micron All Sky Survey (2MASS) catalogue \citep{skru06}. 
The optical and NIR magnitudes of the SN are reported in Tables \ref{tabA1} and \ref{tabA2}, while
those of the local sequence stars are in Table \ref{tabA3}.

The pre-reduction process of optical spectroscopy frames followed the steps described above for photometric frames. Then,
one-dimensional spectra were optimally extracted in IRAF, and wavelength calibrated using reference spectra of arc lamps.
Finally, sensitivity curves obtained using spectra of spectro-photometric standards allowed us to flux-calibrate the SN spectra.
Telluric absorption features were removed using normalized telluric absorption profiles derived from the spectrum of the standard star.

\begin{figure}
\includegraphics[width=3.55in]{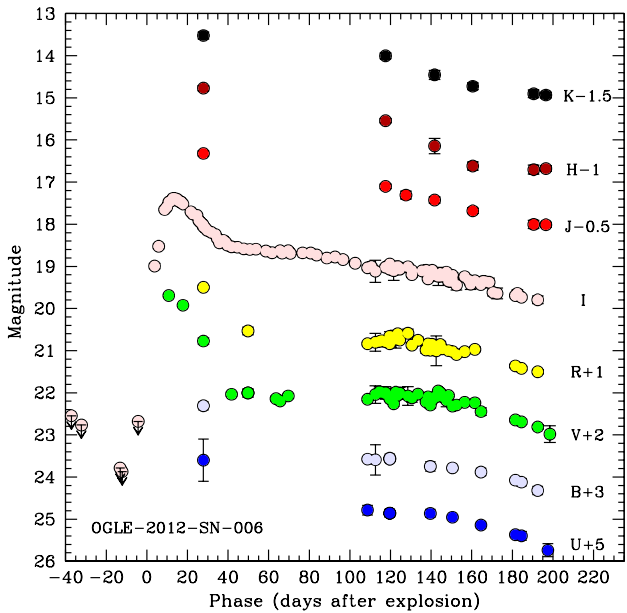}
\caption{Optical and NIR light curves of OGLE-2012-SN-006. The light curves are shifted by the same amount indicated by the labels to the right.
The phase axis is computed in days since explosion ($JD$ = 2456203.8 $\pm$ 4.0).} 
\label{fig2}
\end{figure}

For the NIR spectra, two additional steps were required: firstly, we had to remove the intense background emission.
This was achieved through the subtraction of two consecutive exposures (one from the other),
since during the observation the source was dithered along the slit direction. Secondly, we used spectra of telluric standard stars to
remove the very broad (typically saturated in the NIR) atmospheric absorption bands.

Finally, the consistency of the spectroscopic flux calibration in the optical and NIR spectra was checked with available SN magnitudes 
and, in case of a discrepancy, the spectral fluxes were rescaled to the photometry. The log of spectroscopic observations of OGLE-2012-SN-006 is reported 
in Table \ref{tab1}. The spectroscopic data of OGLE-2012-SN-006 are available from the ESO archive as PESSTO Phase 3 data products
(see {\it http://www.pessto.org}) and are also available via WISeREP \citep{yar12}.\footnote{We note that the data presented and
analysed in this paper were custom reduced, and not obtained from the First Release of $PESSTO$ Spectral data products (SSDR1). This is because the analysis was performed before $PESSTO$ 
SSDR1, and the faintness of the object required a careful treatment. }

\subsection{Analysis of the Light Curves}  \label{lc}

Routine $I-$band observations of the $OGLE-IV$ fields resulted in an excellent monitoring of the region where OGLE-2012-SN-006 exploded. $OGLE-IV$ observations \citep{wyr14}
of the SN field were collected since early July 2012,
and the last non-detection was registered on 2012 September 29.26 UT. The first detection of OGLE-2012-SN-006 was on 2012 October 7.34 UT, so we can constrain the explosion epoch with
a relatively small uncertainty to 2012 October 3.30 UT ($JD$ = 2456203.8 $\pm$ 4.0). After discovery, the object showed a relatively fast light curve rise to $I-$band maximum, 
lasting $\sim$ 2 weeks from the adopted explosion time. 
Using a low-order polynomial fit to the data, we estimated the magnitude of the $I-$band light curve maximum to be $I = 17.41 \pm 0.02$ (on 2012 October 17.6 UT, i.e. $JD$ = 2456218.1 $\pm$ 1.8).
Adopting a distance modulus of $\mu$ = 36.94 $\pm$ 0.19 mag 
for the galaxy hosting OGLE-2012-SN-006 (see Section \ref{intro}), assuming 
negligible host galaxy reddening\footnote{From the earliest spectrum (see Section \ref{spec}), we do not see any clear signature of the narrow interstellar Na I doublet (Na ID) at the parent galaxy redshift, suggesting little (if any)
 contribution of the host galaxy to the total reddening of OGLE-2012-SN-006.} and  a Galactic contribution to the total extinction of $A_I$ = 0.12 mag \citep[obtained adopting the dust map calibration of][]{sch11}, we obtain an absolute
magnitude at peak of $M_I$ = -19.65 $\pm$ 0.19. After maximum, the $I-$band light curve experienced a relatively fast decline (with a slope of $\gamma_1 \approx$ 4 mag 
100d$^{-1}$) lasting about 25 days, followed by a flattening from +25 days to +70 days past maximum ($\gamma_2 \approx$ 0.40 mag 100d$^{-1}$). 
Such an optical/NIR flattening, making the magnitude decline rate slower than expected for a SN powered by radioactivity of $^{56}$Co, has
never been observed in any SN Ibn. After that, the $I$-band light curve becomes slightly steeper ($\gamma_3 \approx$ 0.74 mag 100d$^{-1}$), showing a slope which
is closer to $^{56}$Co decay (0.98 mag 100d$^{-1}$), but still slightly flatter than that slope. A similar decline rate in the late-time luminosity is measured in 
all optical bands. We note that the relatively slow late-time decline of OGLE-2012-SN-006 looks also
alike that observed in some H-rich interacting SNe, such as the Type IIn SN 2010jl (see below).
The increased slope of the $I$-band light curve at very late phases is confirmed by a very late detection on 2013 August 1st,
and by deep detection limits at later epochs. This is possibly an indication of dust formation at very late phases. We note that spectroscopic indicators also support this claim (see Section \ref{spec}).

\begin{figure}
\includegraphics[width=3.55in]{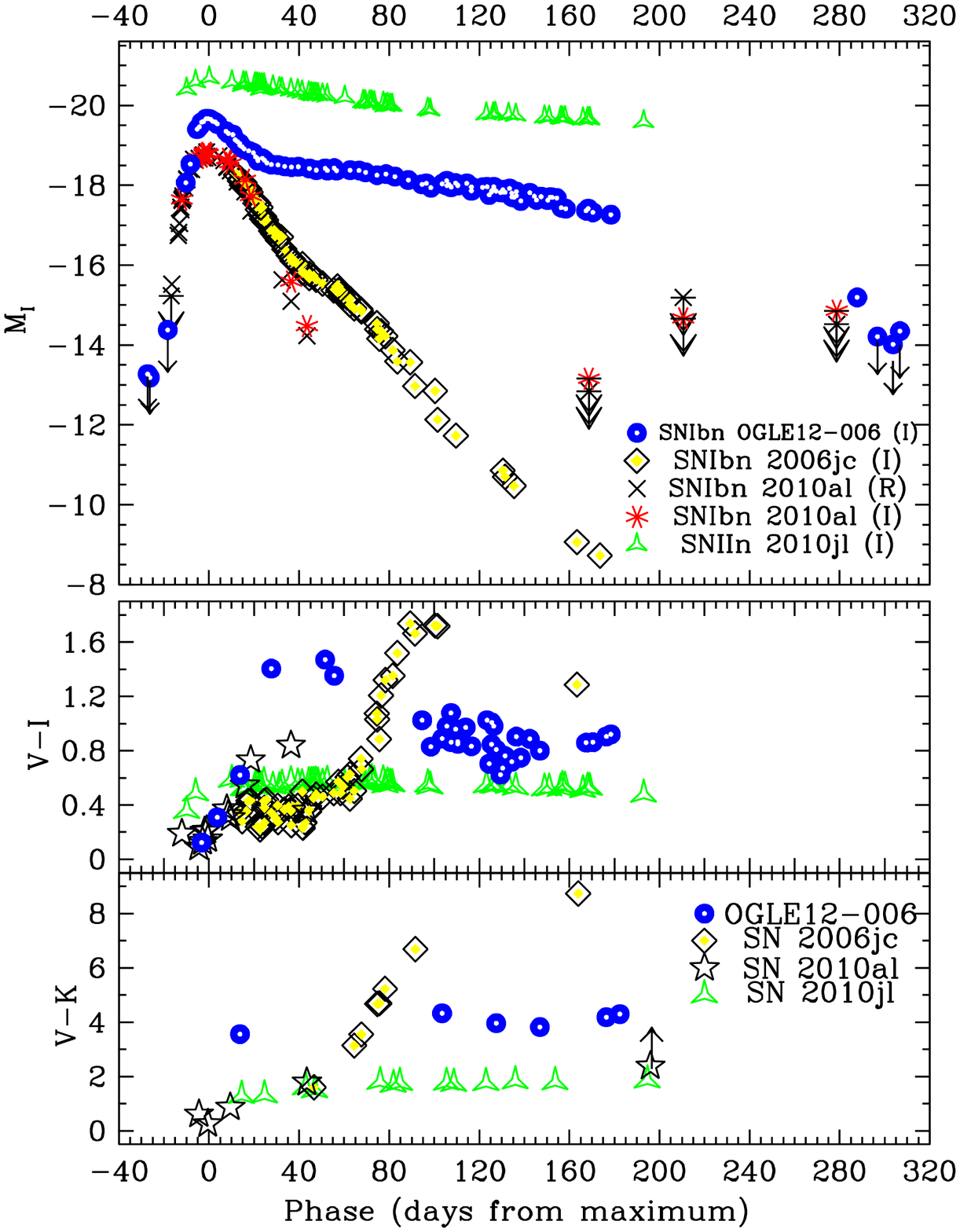}
\caption{Comparisons of the $I$-band absolute light curves (top), $V-I$ (middle) and $V-K$ (bottom) colour curves of SNe OGLE-2012-SN-006, 2010al, 
2006jc and the Type IIn SN 2010jl. Data of SN 2010al are from \protect\citet{pasto13a}, those of SN 2006jc are from 
\protect\citet{pasto07,pasto08a,fol07,seppo08,dic08,anu09} and those of SN 2010jl are from \protect\citet{sto11,zha12,fra14}.
In the top panel, together with the $I$-band light curves of the SN sample, the well-followed 
$R$-band light curve of SN 2010al has been included.} 
\label{fig3}
\end{figure}

\begin{table*}
\caption{Log of the spectroscopic observations of OGLE-2012-SN-006. }
\begin{center}
\small
\begin{tabular}{cccccc} \hline\hline
Date &   JD-2400000  & Days since maximum & Instrumental configuration   & Range (\AA)  & Resolution (\AA) \\ \hline
Oct 15, 2012 & 56215.81 & -2.3 & DuPont+B$\&$C & 3600-10000 & 27 \\
Jan 10, 2013 & 56302.70 & +84.6 & DuPont+WFCCD & 3750-9190 & 8 \\
Jan 20, 2013 & 56312.56 & +94.5 & NTT+EFOSC2+gm11+gm16 & 3360-10090 & 14;14 \\
Feb 20, 2013 & 56343.68 & +125.6 & NTT+EFOSC2+gm13 & 3680-9300 & 18 \\ 
Feb 22, 2013 & 56345.64 & +127.5 & NTT+SOFI+Blue+Red & 9360-25000 & 24;32 \\
Mar 18, 2013 & 56369.64 & +151.5 & NTT+EFOSC2+gm13 & 3650-9300 & 18 \\
Apr 19, 2013 & 56402.49 & +184.4 & NTT+EFOSC2+gm13 & 3660-9280 & 18 \\
\hline
\end{tabular}
\end{center}
\label{tab1}
\end{table*}

\begin{figure}
\includegraphics[scale=.414,angle=0]{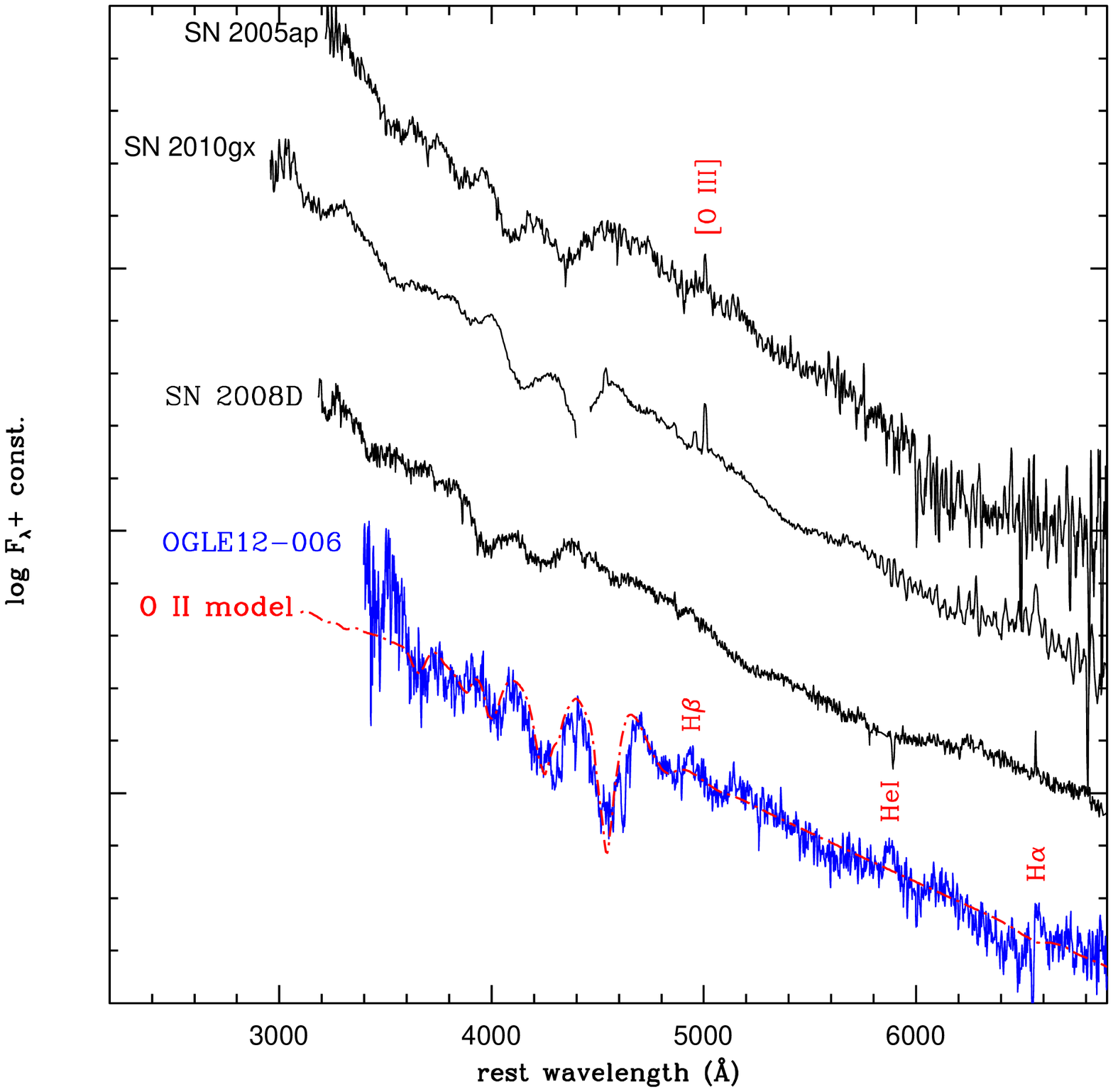}
\caption{Comparison of our earliest spectrum of OGLE-2012-SN-006 (phase = -2.4 days from maximum) with pre-maximum spectra of other stripped envelope SNe.
Our sample includes the super-luminous SNe 2005ap \citep[][phase = -3.6 days from maximum]{qui07} and 2010gx \citep[][time dilation corrected phase = -4.3 days from maximum]{pasto10}, 
and a spectrum of the Type Ib SN 2008D obtained soon after the shock breakout \citep[][i.e. about 1.8 days after the X-ray burst, and -16.6 days from the $V$-band maximum]{mod09}. 
A {\it SYNOW} model, labelled as  ``O II model'', is overplotted to the observed spectrum of OGLE-2012-SN-006. The model
was obtained adopting a continuum temperature of 11000 K, a photospheric velocity of 7000 km s$^{-1}$, and 
including a single contributing species, O II.} 
\label{fig4}
\end{figure}

During early phases, the evolution of OGLE-2012-SN-006 was poorly monitored in the other bands. The object was occasionally observed in the
$V$-band by the $OGLE-IV$ survey, as well as in the $g', r',  i' $ Sloan bands\footnote{Sloan $g', r',  i' $ magnitudes were transformed into
Johnson $B$, $V$, $R$ and $I$ magnitudes following the prescriptions reported in {\it http://www.mpe.mpg.de/$^\sim$jcg/GROND/calibration.html}.} and in the near-infrared (NIR) with the 2.2-m MPI/ESO Telescope in La Silla equipped with GROND \citep{gre08}.
However, after the SN classification, we realized that this was the first opportunity to monitor a Type Ibn SN until very late phases, since the object was showing a very slow photometric evolution.
Therefore we increased the monitoring frequency of the $PESSTO$ campaign in the optical and NIR bands using the NTT equipped
with EFOSC2 and SOFI, and the 1-m LCOGT Telescope located at Cerro Tololo Inter-American Observatory (CTIO). This late-time monitoring allowed 
us to follow in detail the late evolution of OGLE-2012-SN-006, to compute its quasi-bolometric light curve and eventually put an upper limit to the $^{56}$Ni mass.

In Figure \ref{fig3} (top) the $I$-band absolute light curve of OGLE-2012-SN-006 is compared with those of SNe 2006jc, 2010al and 2010jl (the latter being a Type IIn event). 
For SN 2006jc we adopted $\mu$ = 32.01 mag, $A_{I,tot}$ = 0.03 mag and JD(max) = 2454008, for SN 2010al
$\mu$ = 34.25 mag, $A_{I,tot}$ = 0.10 mag and JD(max) = 2455285, while for SN 2010jl $\mu$ = 33.45 mag, $A_{I,tot}$ = 0.09 mag and JD(I,max) = 2455494.
The distance and the reddening adopted for OGLE-2012-SN-006 are those discussed above. For SN 2010al the 
best monitored $R$-band light curve is also shown as a comparison. OGLE-2012-SN-006  is significantly 
brighter than the other two SNe Ibn during its entire evolution  (though fainter than SN 2010jl).
We note that the early post-peak luminosity declines of SNe 2010al and 2006jc are  similar, whilst OGLE-2012-SN-006 flattens quite early (at phase $\sim$ 25-30 days past maximum). 
However, the late-time photometric behaviour of SN 2006jc (and, possibly, SN 2010al) is mostly determined by dust formation which produces a flux deficit in the optical
and an enhanced  emission in the infrared domain \citep{smi08,seppo08,pasto13a}. Clearly, this was not observed in OGLE-2012-SN-006.
We also note that the decline rate of the $I$-band light curve of OGLE-2012-SN-006 at phases later than $\sim$25 days after maximum is similar to that of the the Type IIn SN 2010jl, 
which showed unequivocal evidence of ejecta-CSM interaction at all evolutionary stages.

In the middle panel of Figure \ref{fig3}, the $V-I$ colour curves of the above-mentioned objects are shown. We note that OGLE-2012-SN-006 becomes
red very rapidly, moving from a blue colour at early stages to $V-I \approx 1.4$ mag at about 30 days after maximum. Then the $V-I$ colour turns
slightly bluer in the following couple of months, and thereafter settles to about 0.8 mag. The evolution of the $V-I$ colour in SN 2006jc
is quite different, since it remains almost constant (0.3-0.4 mag) until $\sim$50 days. Later on, it rapidly rises to $V-I \approx$ 1.6 mag up to day 90,
and finally declines to 1.2 mag at about 160 days. Unfortunately, for SN 2010al, the $V-I$ colour monitoring was limited only at the early phases
\citep{pasto13a}. However, from the available data, the evolution of the three SNe is substantially similar until day $\sim$20.

\begin{figure*}
\includegraphics[scale=.75,angle=0]{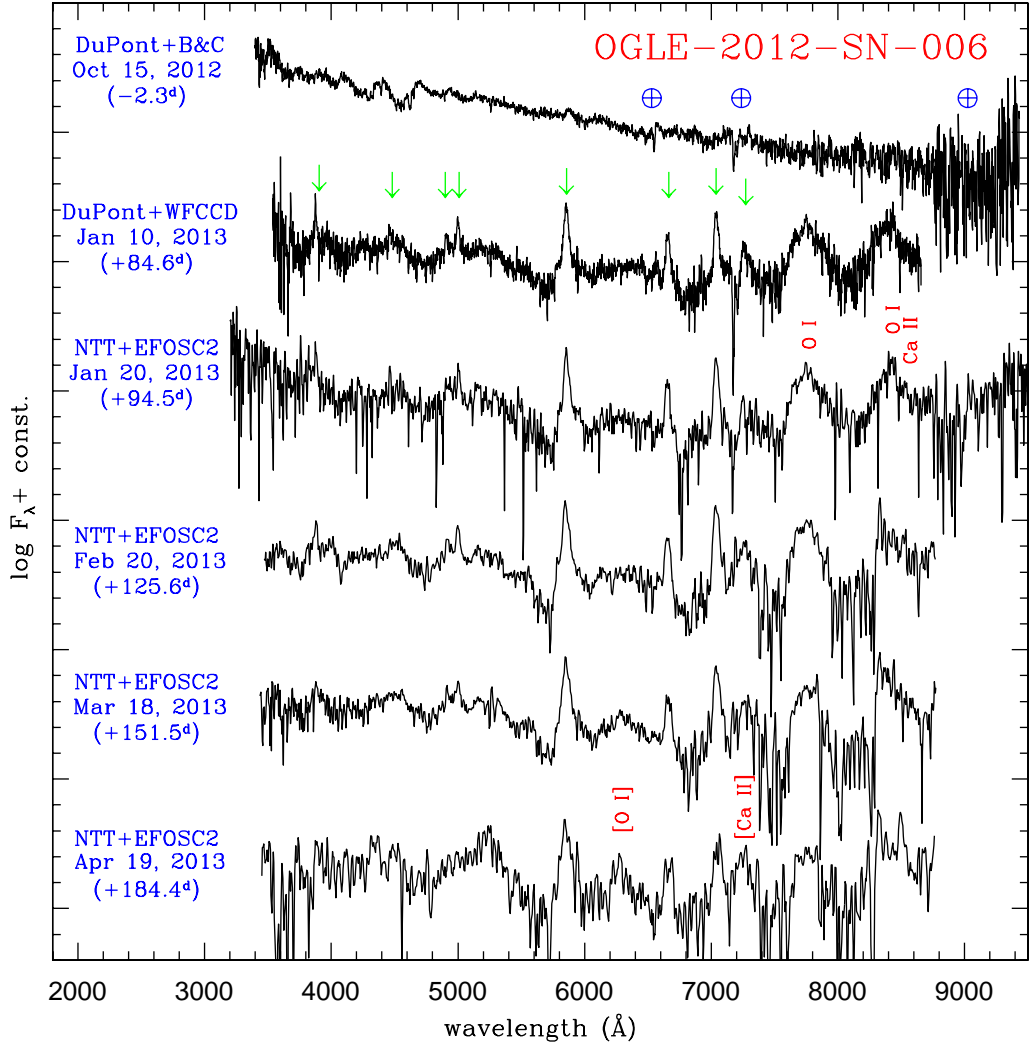}
\caption{Sequence of optical spectra of OGLE-2012-SN-006 reported at the rest wavelength frame. The positions of the strongest He I lines are marked with green arrows.
The most important SN nebular lines are also labelled. The positions of the most prominent telluric features are marked with the ``$\oplus$'' symbol. The phases reported
in parenthesis are days from the maximum light.
} 
\label{fig5}
\end{figure*}

The bottom panel of Figure \ref{fig3} shows the $V-K$ colour evolution. Whilst the $V-K$ colour in OGLE-2012-SN-006 remains roughly constant over the 
entire monitored evolution ($V-K \approx$ 4 mag, ranging from 3.5 mag at early phases to 4.3 mag at late phases), the $V-K$ colour of SN 2006jc shows a monotonic rise
from $\sim$1.4 mag at 1.5 months after maximum to  $\sim$8.8 mag at day 160. Again, the $V-K$ colour of SN 2010al is available only at early phases 
(from around the light curve peak to 1.5 months after), and - in terms of time evolution - apparently mimics that of SN 2006jc. We note that 
colours of the SN IIn 2010jl evolve very little during the first 6 months of its evolution, and remain quite blue.

\subsection{Spectral Evolution}  \label{spec}

The object was discovered at very early stages, and the first spectrum (2012 October 15; Figure \ref{fig4}), dominated by a blue, almost featureless continuum, did not allow a secure classification 
of the object.
The temperature as obtained from a blackbody fit to the spectral continuum is $11000 \pm 1200$ K.
Weak lines of He I  $\lambda$5876, H$\alpha$ and H$\beta$~ are detected. A number of broad absorption features are visible at wavelengths shorter than 5000 \AA.
These absorptions somewhat resemble the O II features observed in spectra of super-luminous stripped envelope SNe \citep[such as SNe 2005ap and 2010gx, see Figure \ref{fig4},][]{qui11,pasto10} 
or even the putative higher ionization CNO lines detected by \citet{mod09} in the earliest spectrum of the more canonical Type Ib SN 2008D. 
In order to identify the lines in this early spectrum, we computed a model using the spectral synthesis code
{\it SYNOW} \citep{fis00}. We adopt a continuum temperature of 11000 K and a
velocity at the photosphere of 7000 km s$^{-1}$, which is significantly lower (by a factor 2 to 3) than those measured in the comparison objects of Figure \ref{fig4}.
The contribution of O II alone is sufficient to reproduce most spectral features   in the wavelengths range between 3500 \AA~and 5000 \AA.
We find a good match between our synthetic {\it SYNOW} spectrum and the observed one, 
making the identification of the spectral features as O II quite robust. However, we cannot rule out a contribution from other CNO ions, including N III, 
to shape the observed features  in the October 15 spectrum of OGLE-2012-SN-006.

\begin{figure}
\includegraphics[scale=.43,angle=0]{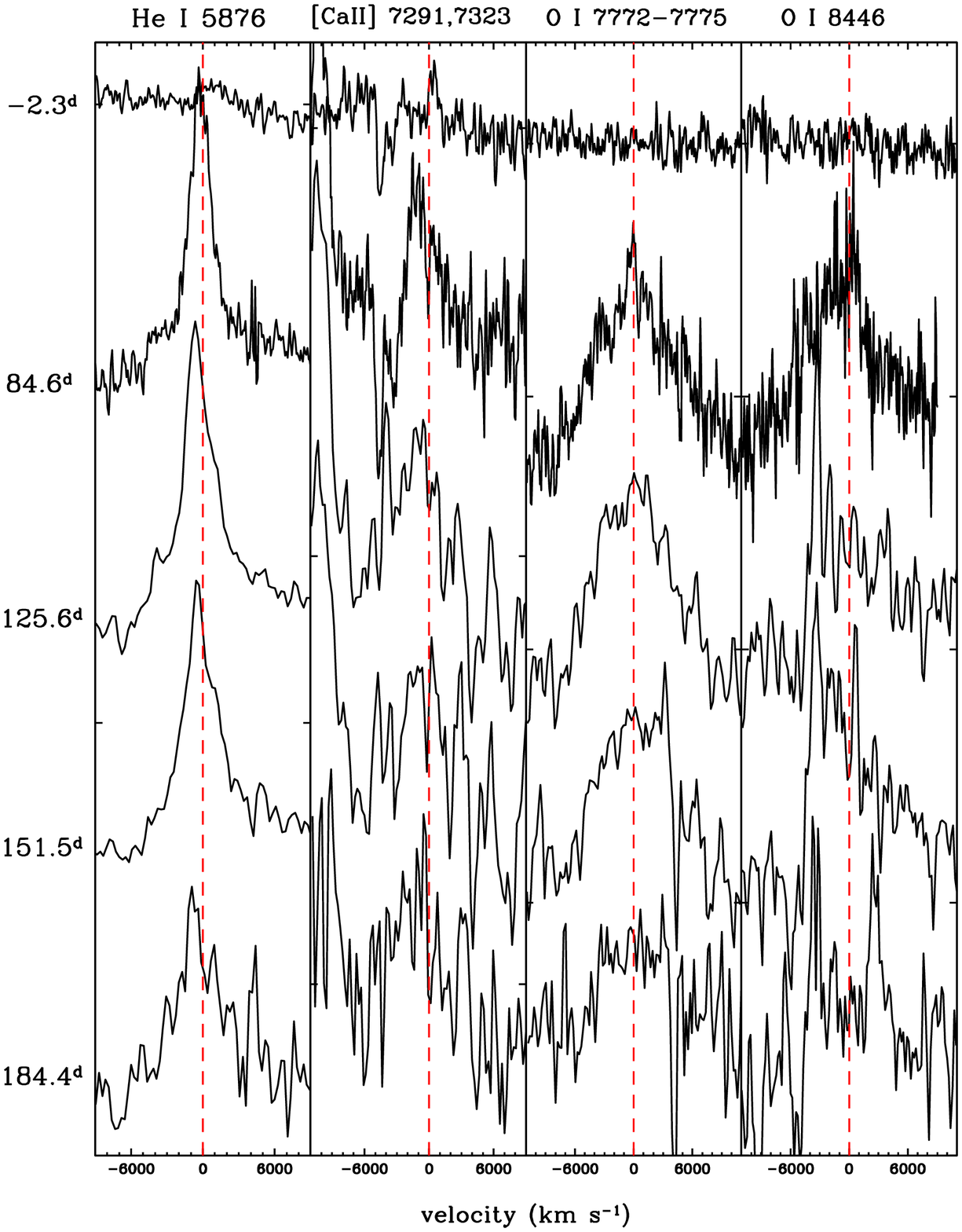}
\caption{ From the left to the right: evolution of the profiles of He I  $\lambda$5876, [Ca II]   $\lambda\lambda$7291,7323, the O I   $\lambda\lambda$7772-7775
and O I  $\lambda$8446. The numbers on the left indicate the phases of the spectra from the maximum light.} 
\label{fig6}
\end{figure}

\begin{figure}
\includegraphics[scale=.43,angle=0]{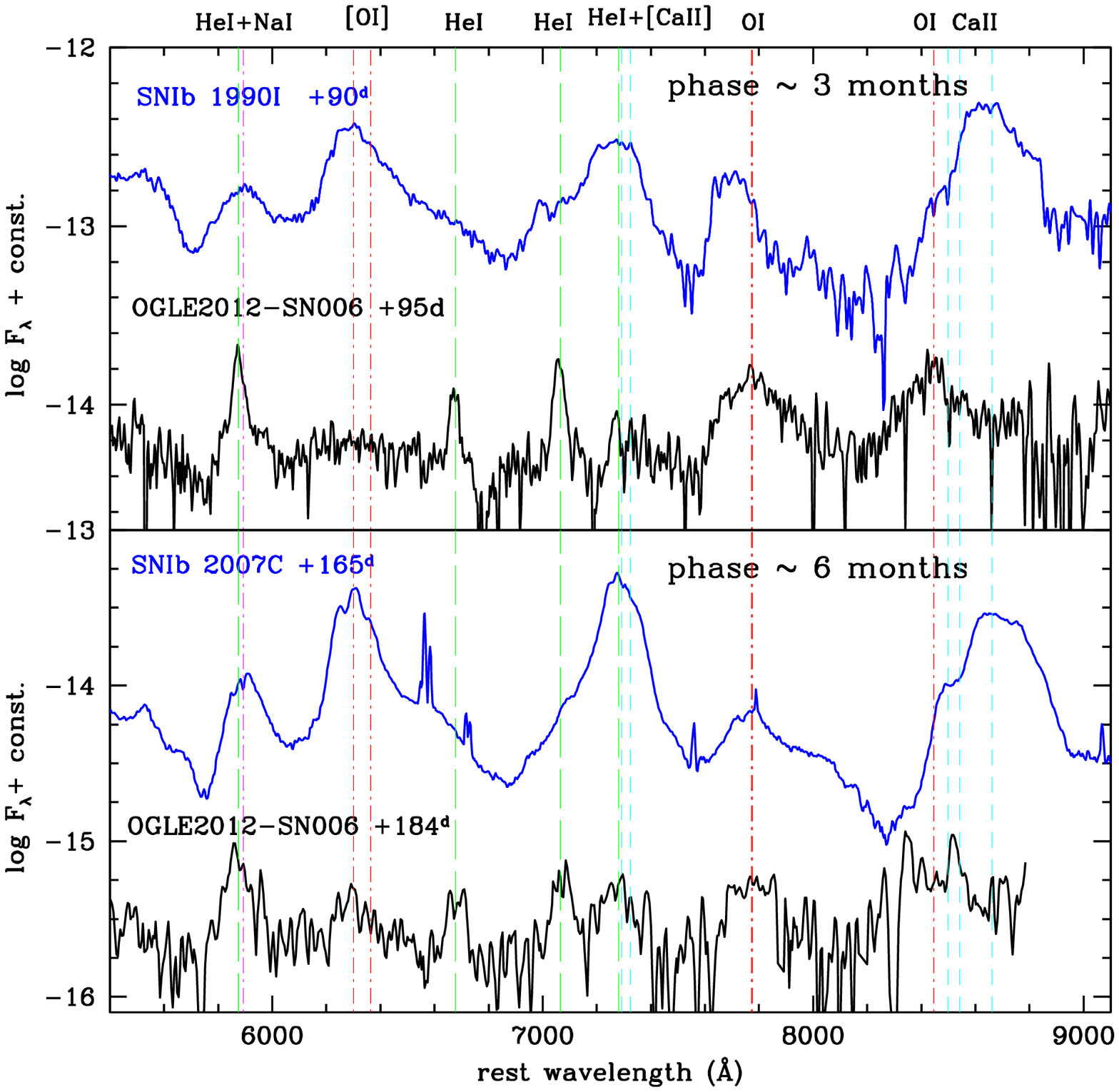}
\caption{Top:  Comparison between spectra of OGLE-2012-SN-006 and the Type Ib SN 1990I at about 3 months after maximum. Bottom: a later ($\sim$6 months) spectrum of  OGLE-2012-SN-006 
is compared with one of the SN Ib 2007C at a similar phase. The spectra of the comparison Type Ib SNe are from \citet{tau13}. The position of the most important lines is marked
with vertical lines: He I (green colour), Na I (magenta), O I (red) and Ca II (cyan).
} 
\label{fig7}
\end{figure}

Another spectrum was obtained on 2013 January 10, and allowed \citet{pri13} to classify the object as a Type Ibn SN because of the presence of strong, narrow He I lines
in emission. These first two spectra of our sequence were both obtained with
the 2.5-m DuPont Telescope at the Las Campanas Observatory (LCO), the former using a B$\&$C spectrograph, the latter with WFCCD. After the classification announcement, i.e. when the object was
$\sim$3 months old, we started a continuous (though relaxed) $PESSTO$ monitoring campaign using the NTT (with EFOSC2 and SOFI). Our complete spectral sequence is shown in Figure \ref{fig5}.

The optical spectra obtained in January and February 2013 display very little evolution.
They show a blue spectral energy distribution (SED), with a clear flux attenuation at wavelengths longer than about 5600 \AA. 
The blue SED is a common property of Type Ibn SNe as well as other interacting events, and is possibly generated by blends of Fe emission lines \citep[e.g.][]{smi09}. 
A very broad absorption feature, with a FWHM $\sim$ 270 \AA, is 
detected at about 4800 \AA. This is a common feature in SNe Ibn, and is probably due to Fe II lines. Much narrower He I lines are clearly detected in emission with a 
$v_{FWHM} \approx$ 1300 km s$^{-1}$. 
The strongest line at optical wavelengths is the He I $\lambda$5876 emission ($v_{FWHM} \approx 2100$ km s$^{-1}$), possibly superposed on a less prominent, broader component  ($v_{FWHM} \approx 7500$ km s$^{-1}$). 

Other prominent and narrow (but resolved) lines are He I $\lambda$6678 and  He I $\lambda$7065 ($v_{FWHM} \approx 1400-1500$ km s$^{-1}$). 
At this time, there is no evidence for the presence of a narrow H$\alpha$.
The He I $\lambda$7281 emission appears to be broader than other He I lines,  having $v_{FWHM} \approx$ 2600 km s$^{-1}$, though its profile may be warped by the telluric absorption band. We may also consider the possibility of a 
blend with nebular emission lines at about 7300 \AA, such as [Ca II]  $\lambda\lambda$7291,7323 or, less likely, [O II]  $\lambda\lambda$7320,7330.
Other weaker He I lines identified in the optical spectra are $\lambda$3889,  $\lambda$4471,  $\lambda$4922 and  $\lambda$5016.
At longer wavelengths, we clearly note the presence of very broad emissions identified as O I at 7772-7775 \AA~  (possibly blended with Mg II, with
$v_{FWHM} \approx 10^4$ km s$^{-1}$), O I $\lambda$8446 and, though weaker, the NIR Ca II triplet.

The two latest spectra (obtained in March and April 2013) show a new feature at about 6300 \AA, that becomes more prominent with time. We identify it as the emerging
[O I]  $\lambda\lambda$6300,6364 doublet typical of core-collapse SNe. In addition, in the 2013 April 19th spectrum, the 7300 \AA~feature has now similar strength as the He I  $\lambda$7065 emission.
This is probably an indication of the increased contribution of the [Ca II]  $\lambda\lambda$7291,7323 (or, perhaps, [O II]  $\lambda\lambda$7320,7330) component in the blend with He I  $\lambda$7281.
In Figure \ref{fig6} we show the time evolution of the profiles of representative spectral lines in OGLE-2012-SN-006, in the velocity space: He I  $\lambda$5876, the [Ca II]  $\lambda\lambda$7291,7323 doublet, the O I features at 7772-7775 \AA~
and 8446 \AA. In Figure \ref{fig7} late-time (3 and 6 months past maximum) spectra of OGLE-2012-SN-006 are compared with those of the Type Ib SNe 1990I and 2007C at 
similar phases. We note than, in contrast with that observed in late Type Ib SNe, in OGLE-2012-SN-006 the O I  $\lambda$8446 line is the most prominent feature at the red wavelengths, and largely
dominates over the Ca II NIR triplet.

\begin{figure*}
\includegraphics[scale=.6,angle=270]{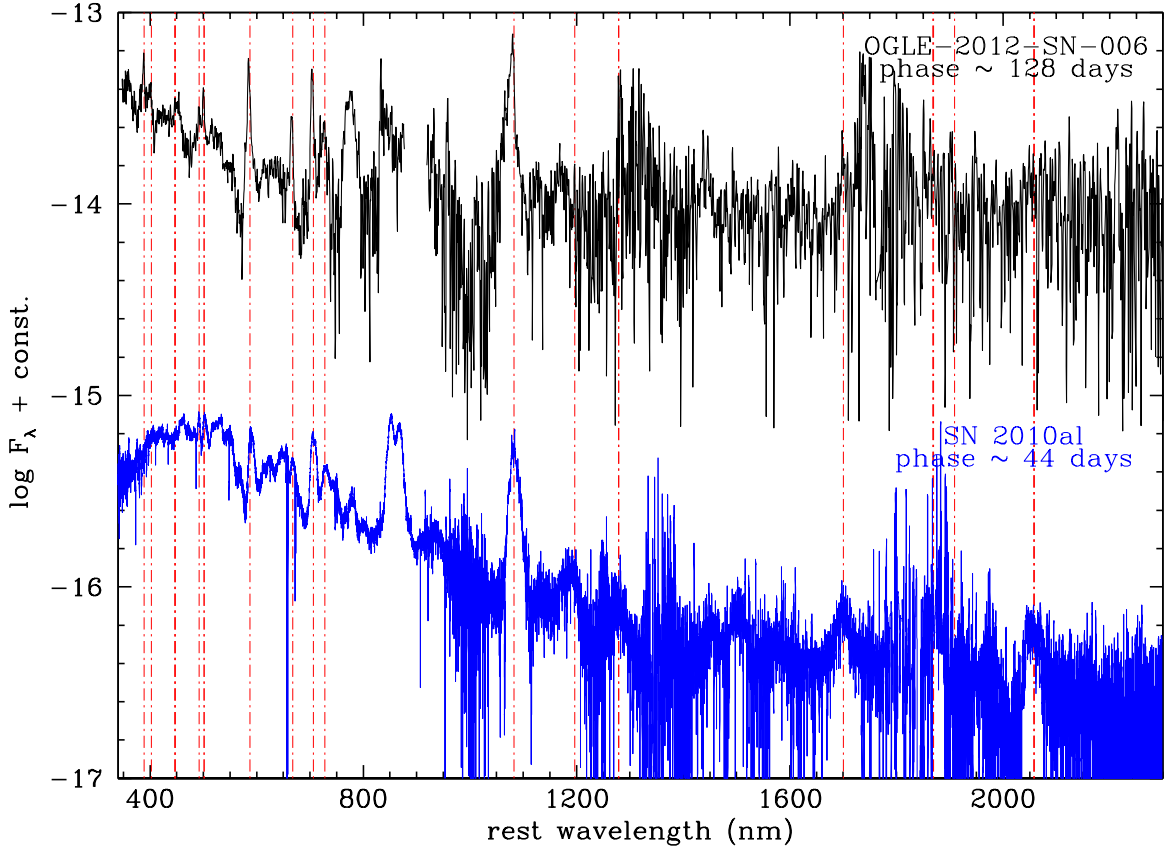}
\caption{Optical + NIR spectrum of OGLE-2012-SN-006 {($\sim$128 days after maximum) compared with an X-Shooter spectrum of SN 2010al \citep{pasto13b}
obtained when the SN was about 44 days past maximum (i.e. 2 months after the explosion)}. The red dot-dashed lines mark the position of the strongest He I features.
The two spectra have been redshift-corrected. Both objects show rather prominent features of O I  $\lambda$7774 (much stronger in OGLE-2012-SN-006), and O I $\lambda$8446 plus the Ca II NIR triplet.} 
\label{fig8}
\end{figure*}

The He I emission lines have a slightly blue-shifted peak, and the amount of blue-shift increases with time, 
from $\sim$ 250 km s$^{-1}$ in the two January 2013 spectra 
to about 400-500 km s$^{-1}$ in the February-March 2013 spectra (see also Figure \ref{fig6}, left panel). 
Unfortunately, the last available spectrum (April 19, 2013) has a very low signal-to-noise. However, from the position of the 
He I  $\lambda$5876 emission, we tentatively infer a blue-shift of 700 $\pm$ 150 km s$^{-1}$. In addition, the emission profiles show an evident asymmetry. All of this 
would support our claim (Section \ref{lc}) that some dust is forming in OGLE-2012-SN-006 at late phases. We remark that in SN 2006jc the formation of dust was observed in a post-shock cool dense shell starting from $\sim$50 days after the SN explosion \citep{smi08,seppo08}. Blue-shifted line emissions are frequently observed in late-time spectra of SNe IIn, and are
usually considered a signature of dust formation \citep[e.g.][]{ger00,fra05,str12}. However, an evident blueshift of the broad H$\alpha$ line component was observed in the Type IIn SN 2010jl \citep[see][their figure 12]{fra14}, and was interpreted as resulting from the bulk gas velocity, likely due to acceleration of unshocked gas
by the SN radiation field.

In Figure \ref{fig8} a late-time optical + NIR spectrum of OGLE-2012-SN-006 obtained with the NTT between 2013 January 19th and 21st (at about 128 days after maximum) is compared with a VLT + XShooter
spectrum of the Type Ibn SN 2010al (phase $\sim$ 44 days since light curve peak) presented in \citet{pasto13a}. Despite the moderate signal-to-noise, the two spectra are very similar, being dominated by prominent
He I lines (marked with red dot-dashed lines in the figure). In particular, the strongest He I line is the $\lambda$10830 feature. Other He I features clearly detected in the NIR spectrum of OGLE-2012-SN-006 are the $\lambda$17002 and $\lambda$20581 lines,
and a blend at around 18690 \AA. We also notice the overall similarity of the two spectra in the optical domain, both showing the prominent blend formed by O I $\lambda$8446 with the NIR Ca II triplet, while the
O I $\lambda$7774 feature is much stronger in the spectrum of OGLE-2012-SN-006 (although it is clearly detected also in SN 2010al). 

\section{Discussion and summary}

OGLE-2012-SN-006 is an interesting object owing to its unprecedented properties in the context of the Type Ibn SN family. An early-time spectrum 
 shows a blue continuum and broad absorptions probably due to CNO elements (mostly O II), that are not frequently observed in classical core-collapse SN spectra. 
These spectral features are similar to those observed in super-luminous stripped-envelope transients \citep{qui11}. The  post-peak optical and NIR light 
curves show a fast decline until about one month after the explosion. Then the light curves level off to a very slow decline rate. 
In coincidence with the late curve flattening, the spectra show the relatively narrow He I in emission typical of the Type Ibn SN family,
which are interpreted in terms of interaction with He-rich CSM.
The late light curve tail has initially a flat slope, although later (starting from about 4 months after the explosion) it shows a decline rate similar to that 
expected when the powering mechanism is 
the radioactive decay of $^{56}$Co into $^{56}$Fe. The optical light curves of the prototype SN 2006jc showed a much faster decline in the optical bands and
an increasing luminosity in the IR domain. This was interpreted as a signature of dust formation \citep{smi08,seppo08,dic08,noz08,sak09} in a post-shock
cool and dense circumstellar shell. However, there is evidence of an increasing slope in the late-time $I$-band light curve in OGLE-2012-SN-006, which may be
attributed to dust formation also in this case.

\begin{figure}
\includegraphics[scale=.435,angle=0]{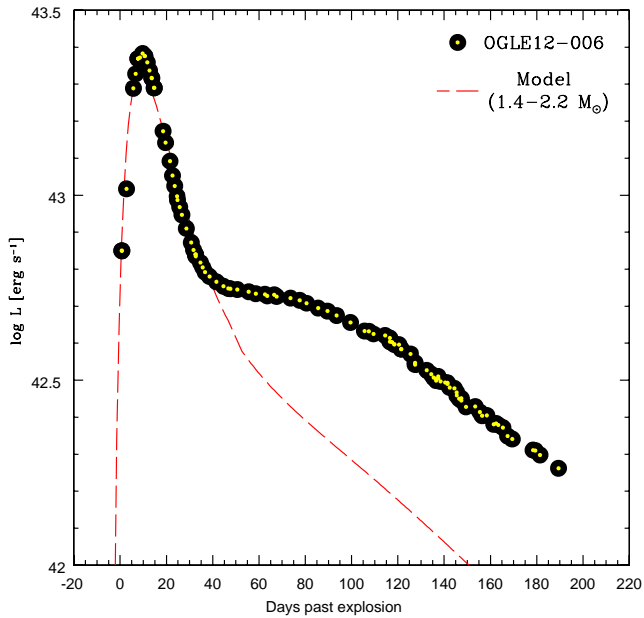}
\caption{Comparison between the pseudo-bolometric light curve of OGLE-2012-SN-006 and a radioactively-powered light curve model. As we have limited information on the ejecta 
velocity at the time of the light curve peak, we assume that the photospheric velocity at maximum is v = 15000 $\pm$ 1000 km s$^{-1}$.  We adopt an optical opacity of 
$\chi$ = 0.1 cm$^2$ g$^{-1}$. A model which provides a decent match
with the observed quasi-bolometric light curve of OGLE-2012-SN-006 at early phases, is obtained adopting the following parameters: 
M($^{56}$Ni) = 0.8 - 1.1 M$_\odot$, M$_{ej}$ = 1.4 - 2.2 M$_\odot$, E$_K = 1.3 - 7.5 \times 10^{51}$ erg. Clearly, a poor match is obtained with the late-time
light curve of OGLE-2012-SN-006.
} 
\label{fig9}
\end{figure}

An important issue that we have not yet discussed is the physical mechanisms that could power the bright peak luminosity and the unprecedented photometric evolution of
the Type Ibn OGLE-2012-SN-006. For this goal, we have calculated a (quasi)-bolometric light curve for OGLE-2012-SN-006 by 
 integrating the flux contribution in individual optical and NIR pass-bands.
 In practice, for each epoch with $I$-band observations available and for each band, we derived the flux at the effective wavelength.
When photometric points in individual bands were not available at a given epoch, their contribution was estimated through an interpolation of
 photometry in adjacent epochs. Unfortunately, early-time photometry was available for OGLE-2012-SN-006 only in the $I$ and (to a lesser extent) in the $V$ band.
The flux contribution of the missing bands at early phases, in particular during the rising branch to the peak and the period around maximum,
has been obtained under the rough assumption that the early colour evolution of OGLE-2012-SN-006 was similar to that of the Type Ibn SN  2010al \citep{pasto13a}.
Although the colour evolutions of SNe 2010al and OGLE-2012-SN-006 are very different, there is a good match in the early-time
optical colours (up to $\sim$15 days past maximum; see Figure \ref{fig3}, middle panel).
The fluxes, corrected for the adopted extinction, provide the spectral energy distribution at the given            
phase, which are then integrated through the trapezoidal rule. The observed flux is finally converted to luminosity      
adopting the distance and interstellar reddening values mentioned above. We did not account for flux contribution beyond the observed         
$U$ and NIR bands, and therefore this has been more properly quoted as a "quasi-bolometric" light curve.  
We should emphasize that the luminosity contribution of the UV domain, especially at early phases, may be significant \citep[see e.g.][]{pasto13a}.
The resulting quasi-bolometric light curve is shown in Figure \ref{fig9}.

Although we have limited information on the spectroscopic and photometric evolution of OGLE-2012-SN-006,
we attempt to constrain SN parameters using the quasi-bolometric light curve computed before,
with the initial assumption that the SN luminosity peak is mostly powered by the radioactive decays.
To this aim, we fit the SN data using a toy light-curve model elaborated by \citet{val08},\footnote{This model is obtained using the code of \citet{val08}, which is based on \citet{arn82},
but accounting for the typo correction reported in \citet{arn96} - \citep[see][]{whee14}.} which has been shown to successfully reproduce
the early-phase bolometric light curves of various stripped-envelope SNe.
The model of \citet{val08} is based on simple analytical approximations  \citep{arn82,cap97,clo97},
and on a division of the light curve into two periods. In the first period, the supernova is optically thick (photospheric phase); later on, it gets optically
thin (nebular phase). The model assumes that - at least in the early phase ($<$ 50 days) - the CSM-ejecta interaction does not provide a major contribution
to the observed bolometric luminosity. By assessing the model consistency, we will make an effort to test this assumption, although it may
appear natural since the early light curve of OGLE-2012-SN-006 resembles those of radioactively-powered Type Ia and/or Type Ib/c SNe, 
and our earliest spectrum shows no narrow emission lines from interaction.
The comparison between the bolometric light curve of OGLE-2012-SN-006 and our best light curve model of a radioactively powered stripped-envelope SN 
is shown in Figure  \ref{fig9} (the parameters used for the model are illustrated in the figure caption).
If the SN radiation is powered by radioactive decays, this would imply a relatively high kinetic energy
and extremely small ejecta mass, a large fraction of which would be $^{56}$Ni. The inferred M($^{56}$Ni)/M$_{ej}$ ratio ($\sim$0.5) is similar to those estimated in some 
luminous Type Ia SNe \citep[e.g.][]{tau13}, but are not comfortably accomodated in the context of a core-collapse explosion scenario. We also note the late-time spectra
are dominated by He and $\alpha$-elements, and not by iron-peak elements. All of this clearly argues against the initial assumptions adopted for the model, and 
suggests that the $^{56}$Ni to $^{56}$Co to $^{56}$Fe decay chain is not the primary process which powers the light curve of OGLE-2012-SN-006. As a consequence, ejecta-CSM
interaction provides a more plausible explanation.

We note, however, that a degenerate progenitor system scenario had been previously proposed by \citet{san13} for another Type Ibn SN, PS1-12sk.
The unusual location of that object in the outskirts of an elliptical galaxy, suggested
a progenitor scenario which could have been different from the canonical core-collapse of a massive, stripped-envelope 
star normally assumed for SNe Ibn, including the possibility of a white dwarf explosion  in a He-rich environment \citep{san13}. 
However, to our knowledge, this still remains a unique case in the Type Ibn SN zoo, since all other objects exploded in spiral galaxies \citep{pasto13b}. 
In addition, as admitted by \citet{san13}, some residual star formation at the SN location and/or the association with a nearby, low 
luminosity dwarf galaxy cannot be ruled out, and -consequently- the progenitor may still be a massive star.

The fact that SNe Ibn explode preferentially in young stellar population environments, the pre-SN eruption detected in SN 2006jc and -even more- 
the detection of the [O I] $\lambda\lambda$ 6300,6364 doublet in the late spectra of OGLE-2012-SN-006, are robust arguments to support a core-collapse scenario for
Type Ibn SNe. As mentioned before, the high M($^{56}$Ni)/M$_{ej}$ ratio estimated for OGLE-2012-SN-006 from the model described above, can be explained with
incorrect initial assumptions on the source powering the early light curve of this object. In other words, the high 
luminosity at peak may have not been completely due to radioactive decay. In the view of the presence of a dense
He shell, we argue that other processes may give a significant contribution to the total SN
luminosity around maximum light. As a consequence, the assumptions from using the toy model are violated, and
the progenitor and explosion parameters are likely to be significantly different from those derived in the simple model fits.
Almost certainly the explosion of OGLE-2012-SN-006 produced a smaller amount of $^{56}$Ni and a larger ejecta mass.

In this context, there is an appealing qualitative similarity between the quasi-bolometric light curve of OGLE-2012-SN-006 
and the light curve models discussed in \citet{falk77}. In particular, there is a surprising match with the light curve models obtained
considering a stellar ejected envelope plus an overlying circumstellar shell \citep[Models B of][see e.g. their figure 10]{falk77}.
According to these models, the luminous peak would be the result of radiative heating and diffusion of radiation ahead of the shock front
in a low-density circumstellar shell. Modern simulations of this class of interaction-powered SNe have been presented in detail, 
for example, by \citet{mor11}. According to \citet{mor12} the absence of prominent narrow lines, as in the case of the early spectrum
of OGLE-2012-SN-006, would not be an indication of lack of interaction. Instead, it could be related with a non-steady mass loss of the 
progenitor prior to the explosion. Although this scenario was used to explain the early behavior of H-rich SNe, it might be extended to 
Type I events, such as OGLE-2012-SN-006. However, while in the case of the models of \citet{falk77} or \citet{mor11} the H envelope 
recombination sets the post-peak luminosity evolution, in SNe Ibn the ejecta-CSM interaction is probably the preponderant
source of energy powering the light curve until late phases.

The precursors of Type Ibn SNe have been proposed to be H-depleted, He-rich  \citep[i.e. ``WN'' type,][]{fol07} Wolf-Rayet (WR) stars or more 
He-poor \cite[of ``WC-O'' type,][]{pasto07,tom08} WR stars. In addition, a few rare and moderately H-rich Ibn events, such as SNe 2005la and 2011hw, 
have been proposed to be produced by stars that
were transiting from luminous blue variable to WR stages \cite[e.g. WN-type,][]{smi12,pasto13a}. In this context, the spectra of OGLE-2012-SN-006 show 
very little evidence for the presence of H lines (and only in the first spectrum), thus suggesting that the precursor was an almost totally H-depleted 
WR star that suffered major mass loss processes before exploding as a core-collapse SN.

So far, we have collected incomplete information from an increasing number of Type Ibn SNe. In order to increase our knowledge on this SN family, 
we need to promptly detect a young SN Ibn just after explosion, and follow it until the nebular phase. With OGLE-2012-SN-006, we partly reached that aim, 
since we collected valuable information on the evolution of a SN Ibn until very late phases.
However, because of the slow reaction of the astronomical community and the unusual properties of the early-time spectrum, it
was not promptly classified. As a consequence, we missed the opportunity  to better sample the impressive spectral metamorphosis  
at early times that can be only inferred from the available data. In addition, we did not target OGLE-2012-SN-006 in the ultra-violet (UV) domain,
which is crucial to properly estimate the UV contribution to the total luminosity during the first few days after the explosion.  
An efficient cooperation and exchange of information within the astronomical community is essential to achieve these goals,   
and we aim to do this as soon as a new suitable young SN Ibn will become available.

\section*{Acknowledgments}

We would like to thank the anonymous referee for providing us with constructive comments and suggestions.

AP, EC, SB and MT are partially supported by the PRIN-INAF 2011 with the project {\it ''Transient Universe: 
from ESO Large to PESSTO''}. 
NER acknowledges the support from the European Union Seventh Framework Programme (FP7/2007-2013) under grant agreement n° 267251 “Astronomy Fellowships in Italy” (AstroFIt).
AMG acknowledges financial support by the MICINN grant AYA2011-24704/ESP, by the ESF EUROCORES Program EuroGENESIS (MINECO grants EUI2009-04170), SGR grants of the Generalitat 
de Catalunya and by the EU-FEDER funds. 
SS acknowledges support from CONICYT-Chile FONDECYT 3140534, Basal-CATA PFB-06/2007, and Project IC120009 "Millennium
Institute of Astrophysics (MAS)" of Iniciativa Cient\'{\i}fica Milenio del Ministerio de Econom\'{\i}a, Fomento y Turismo.

Research leading to these results has received funding
from the European Research Council under the European
Union s Seventh Framework Programme (FP7/2007-2013)/ERC
Grant agreement no. [291222] (PI: SJS) and EU/FP7-ERC grant
no. [307260] (PI: AG-Y). AG-Y is also supported by  The Quantum
Universee I-Core programme by the Israeli Committee for planning
and funding and the ISF, a GIF grant, and the Kimmel award.
The OGLE project has received funding from the European Research Council
under the European Community’s Seventh Framework Programme (FP7/2007-
2013)/ERC grant agreement no. 246678 to AU.
This work is partially supported by the Polish Ministry of Science and Higher Education
 program ''Ideas Plus'' No. IdP2012 000162, and
is also partly supported by the European Union FP7 programme
through ERC grant number 320360

This paper is partially based on observations obtained under the ESO-NTT Large Program 188.D-3003
(the Public ESO Spectroscopic Survey for Transient Objects - {\it PESSTO}).
This work makes use of observations from the LCOGT network, the 1.3m Warsaw university Telescope and the 2.5m du Pont Telescope
at the Las Campanas Observatory, the 2.2m MPI/ESO Telescope and the 3.58m ESO New Technology Telescope at ESO-La Silla.

We thank Jonathan Elliott (MPE Garching) for executing the GROND observation.
AP thanks M. L. Pumo for useful discussions.
Part of the funding for GROND (both hardware as well as personnel) was
generously granted from the Leibniz-Prize to Prof. G. Hasinger (DFG grant HA 1850/28-1).

This publication makes use of data products from the Two Micron All Sky Survey, which is a joint project of the University of Massachusetts and 
the Infrared Processing and Analysis Center/California Institute of Technology, funded by the National Aeronautics and Space Administration and 
the National Science Foundation.
This research has made use of the NASA/IPAC Extragalactic Database (NED) which is operated by the Jet Propulsion 
Laboratory, California Institute of Technology,  under contract with the National Aeronautics and Space Administration.
 We acknowledge the usage of the HyperLeda database (http://leda.univ-lyon1.fr).

%{\it Facilities:} \facility{New Technology Telescope}, \facility{Liverpool Telescope}, \facility{GranTeCan (OSIRIS)}.
%\twocolumn

\appendix

\section[]{Photometry of OGLE-2012-SN-006}

\onecolumn
\begin{longtable}{cccccccc}
\caption{\label{tabA1} Optical photometry of OGLE-2012-SN-006.} \\ \hline \hline
Date & JD & $U$ & $B$ & $V$ & $R$ & $I$ & Instrument \\
(dd/mm/yy) & (+2400000) & & & & & & \\ \hline
 03/07/12 & 56111.92 &      -  &     -  &     -  &     -  &     $>$ 21.16   & 1 \\
 16/07/12 & 56124.93 &      -  &     -  &     -  &     -  &     $>$ 22.29   & 1 \\
 11/08/12 & 56150.88 &      -  &     -  &     -  &     -  &     $>$ 22.93   & 1 \\
 20/08/12 & 56159.91 &      -  &     -  &     -  &     -  &     $>$ 22.97   & 1 \\
 27/08/12 & 56166.87 &      -  &     -  &     -  &     -  &     $>$ 22.55   & 1 \\
 01/09/12 & 56171.84 &      -  &     -  &     -  &     -  &     $>$ 22.77   & 1 \\
 20/09/12 & 56190.88 &      -  &     -  &     -  &     -  &     $>$ 23.79   & 1 \\
 21/09/12 & 56191.86 &      -  &     -  &     -  &     -  &     $>$ 23.87   & 1 \\
 29/09/12 & 56199.76 &      -  &     -  &     -  &     -  &     $>$ 22.68   & 1 \\ 
 07/10/12 & 56207.84 &      -  &     -  &     -  &     -  &     18.99 (0.06) & 1 \\
 09/10/12 & 56209.84 &      -  &     -  &     -  &     -  &     18.52 (0.06) & 1 \\
 12/10/12 & 56212.81 &      -  &     -  &     -  &     -  &     17.65 (0.02) & 1 \\
 13/10/12 & 56213.83 &      -  &     -  &     -  &     -  &     17.59 (0.03) & 1 \\
 14/10/12 & 56214.82 &      -  &     -  &   17.69 (0.03) &     -  &     17.47 (0.02) & 1 \\ 
 15/10/12 & 56215.69 &      -  &     -  &     -  &     -  &     17.46 (0.02) & 1 \\ 
 16/10/12 & 56216.81 &      -  &     -  &     -  &     -  &     17.39 (0.02) & 1 \\ 
 17/10/12 & 56217.75 &      -  &     -  &     -  &     -  &     17.39 (0.02) & 1 \\ 
 18/10/12 & 56218.85 &      -  &     -  &     -  &     -  &     17.41 (0.02) & 1 \\ 
 19/10/12 & 56219.76 &      -  &     -  &     -  &     -  &     17.45 (0.02) & 1 \\
 20/10/12 & 56220.70 &      -  &     -  &     -  &     -  &     17.47 (0.02) & 1 \\ 
 20/10/12 & 56220.82 &      -  &     -  &     -  &     -  &     17.47 (0.02) & 1 \\ 
 21/10/12 & 56221.80 &      -  &     -  &   17.92 (0.03) &     -  &     17.52 (0.02) & 1 \\  
 25/10/12 & 56225.69 &      -  &     -  &     -  &     -  &     17.70 (0.02) & 1 \\ 
 26/10/12 & 56226.70 &      -  &     -  &     -  &     -  &     17.76 (0.03) & 1 \\ 
 28/10/12 & 56228.66 &      -  &     -  &     -  &     -  &     17.79 (0.04) & 1 \\ 
 29/10/12 & 56229.70 &      -  &     -  &     -  &     -  &     17.92 (0.03) & 1 \\ 
 30/10/12 & 56230.69 &      -  &     -  &     -  &     -  &     17.96 (0.03) & 1 \\ 
 31/10/12 & 56231.71 &      -  &     -  &     -  &     -  &     18.01 (0.03) & 1 \\ 
 31/10/12 & 56231.87 &      -  &   19.31 (0.10) &   18.77 (0.08) &  18.50  (0.04) & 18.06 (0.06) & 2 \\ %  GROND 
 01/11/12 & 56232.80 &      -  &     -  &     -  &     -  &     18.09 (0.03) & 1 \\   
 02/11/12 & 56233.77 &      -  &     -  &     -  &     -  &     18.15 (0.03) & 1 \\   
 04/11/12 & 56235.75 &      -  &     -  &     -  &     -  &     18.19 (0.03) & 1 \\   
 06/11/12 & 56237.82 &      -  &     -  &     -  &     -  &     18.24 (0.03) & 1 \\   
 07/11/12 & 56238.79 &      -  &     -  &     -  &     -  &     18.34 (0.03) & 1 \\   
 08/11/12 & 56239.69 &      -  &     -  &     -  &     -  &     18.39 (0.03) & 1 \\   
 08/11/12 & 56239.78 &      -  &     -  &     -  &     -  &     18.44 (0.03) & 1 \\   
 10/11/12 & 56241.77 &      -  &     -  &     -  &     -  &     18.38 (0.04) & 1 \\   
 11/11/12 & 56242.78 &      -  &     -  &     -  &     -  &     18.43 (0.03) & 1 \\   
 12/11/12 & 56243.79 &      -  &     -  &     -  &     -  &     18.50 (0.03) & 1 \\   
 14/11/12 & 56245.77 &      -  &     -  &   20.04 (0.05) &     -  &     18.54 (0.03) & 1 \\ 
 17/11/12 & 56248.70 &      -  &     -  &     -  &     -  &     18.55 (0.03) & 1 \\   
 20/11/12 & 56251.72 &      -  &     -  &     -  &     -  &     18.59 (0.03) & 1 \\   
 22/11/12 & 56253.82 &      -  &     -  &   20.01 (0.10) &  19.53 (0.09) &           -       & 3 \\  % NTT
 22/11/12 & 56253.83 &      -  &     -  &   20.00 (0.06) &     -  &           -       & 3 \\  % NTT 
 23/11/12 & 56254.72 &      -  &     -  &     -  &     -  &     18.60 (0.03) & 1 \\   
 26/11/12 & 56257.73 &      -  &     -  &     -  &     -  &     18.59 (0.04) & 1 \\   
 01/12/12 & 56262.65 &      -  &     -  &     -  &     -  &     18.64 (0.04) & 1 \\   
 04/12/12 & 56265.71 &      -  &     -  &     -  &     -  &     18.68 (0.03) & 1 \\   
 06/12/12 & 56267.65 &      -  &     -  &   20.14 (0.06) &     -  &           -   & 1 \\
 08/12/12 & 56269.67 &      -  &     -  &   20.20 (0.05) &     -  &     18.63 (0.03) & 1 \\
 09/12/12 & 56270.65 &      -  &     -  &     -  &     -  &     18.69 (0.03) &  1 \\ 
 12/12/12 & 56273.67 &      -  &     -  &   20.08 (0.06) &     -  &     18.63 (0.03) & 1 \\ 
 13/12/12 & 56274.71 &      -  &     -  &     -  &     -  &     18.69 (0.03) & 1 \\  
 19/12/12 & 56280.72 &      -  &     -  &     -  &     -  &     18.68 (0.04) & 1 \\  
 23/12/12 & 56284.69 &      -  &     -  &     -  &     -  &     18.69 (0.04) & 1 \\  
 26/12/12 & 56287.64 &      -  &     -  &     -  &     -  &     18.73 (0.04) & 1 \\  
 31/12/12 & 56292.65 &      -  &     -  &     -  &     -  &     18.80 (0.04) & 1 \\     \hline
\\
\caption{continued.} \\
\hline\hline
Date & JD & $U$ & $B$ & $V$ & $R$ & $I$ & Instrument \\
(dd/mm/yy) & (+2400000) & & & & & & \\ \hline
 04/01/13 & 56296.69 &      -  &     -  &     -  &     -  &     18.78 (0.04) & 1 \\   
 08/01/13 & 56300.63 &      -  &     -  &     -  &     -  &     18.84 (0.04) & 1 \\  
 14/01/13 & 56306.62 &      -  &     -  &     -  &     -  &     18.93 (0.04) & 1 \\  
 20/01/13 & 56312.70 &   19.79 (0.11) &  20.58 (0.03) &  20.15 (0.03) &  19.84 (0.03) &     19.03 (0.05) & 3 \\  % NTT
 22/01/13 & 56314.59 &      -  &     -  &     -  &     -  &     19.01 (0.05) & 1 \\  
 24/01/13 & 56316.59 &      -  &  20.59  (0.36) &  20.04  (0.21) &  19.80 (0.21) &     19.12 (0.26) & 4 \\ % LCOGT-AP
 26/01/13 & 56318.59 &      -  &     -  &  19.98  (0.12) &  19.78 (0.11) &           -        &  4 \\ %    LCOGT-AP
 27/01/13 & 56319.59 &      -  &     -  &  20.00  (0.08) &  19.76 (0.09) &           -        &  4 \\ %    LCOGT-AP
 29/01/13 & 56321.59 &      -  &     -  &     -  &     -  &     19.02 (0.05) & 1 \\ 
 29/01/13 & 56321.59 &      -  &     -  &  20.02  (0.12) &  19.79 (0.11) &           -        &  4 \\ %    LCOGT-AP
 31/01/13 & 56323.59 &      -  &     -  &  20.11  (0.05) &  19.66 (0.11) &           -        &  4 \\ %    LCOGT-AP
 31/01/13 & 56323.62 &      -  &     -  &     -  &     -  &     18.94 (0.04) & 1 \\ 
 31/01/13 & 56323.67 &   19.86 (0.09) &  20.55 (0.10) &  20.15  (0.03) &  19.83 (0.06) &     19.07 (0.04) & 3 \\ % NTT 
 31/01/13 & 56323.68 &   19.86 (0.09) &  20.58 (0.05) &     -  &     -  &           -        & 3 \\ % NTT
 01/02/13 & 56324.58 &      -  &     -  &     -  &     -  &     19.03 (0.05) & 1 \\  
 02/02/13 & 56325.55 &      -  &     -  &  20.26 (0.05) &     -  &     19.09 (0.04) & 1 \\  
 02/02/13 & 56325.60 &      -  &     -  &  20.06 (0.06) &  19.70 (0.08) &     19.10 (0.23) &  4 \\ % LCOGT-AP 
 03/02/13 & 56326.60 &      -  &     -  &  19.99 (0.10) &  19.73 (0.21) &           -      &  4 \\ % LCOGT-AP
 04/02/13 & 56327.58 &      -  &     -  &  20.06 (0.07) &  19.60 (0.07) &     19.00 (0.10) &  4 \\ % LCOGT
 05/02/13 & 56328.55 &      -  &     -  &  20.01 (0.06) &     -  &     19.05 (0.05) & 1 \\ 
 05/02/13 & 56328.58 &      -  &     -  &  20.01 (0.15) &  19.75 (0.06) &     19.07 (0.06) &  4 \\ % LCOGT 
 09/02/13 & 56332.58 &      -  &     -  &  20.08 (0.22) &  19.59 (0.10) &     19.01 (0.06) &  4 \\ % LCOGT  
 11/02/13 & 56334.58 &      -  &     -  &     -  &     -  &     19.10 (0.05) & 1 \\  
 11/02/13 & 56334.58 &      -  &     -  &  20.12 (0.10) &  19.87 (0.05) &     19.19 (0.06) &  4 \\ % LCOGT
 14/02/13 & 56337.54 &      -  &     -  &  20.03 (0.05) &  19.75 (0.08) &           -      &  4 \\ % LCOGT-Ste
 16/02/13 & 56339.55 &      -  &     -  &     -  &     -  &     19.11 (0.05) & 1 \\
 18/02/13 & 56341.54 &      -  &     -  &  20.22 (0.01) &  19.98 (0.08) &     19.10 (0.10) &  4 \\ % LCOGT-Ste
 19/02/13 & 56342.53 &      -  &     -  &  20.10 (0.09) &  19.84 (0.01) &     19.30 (0.04) &  4 \\ % LCOGT-Ste
 20/02/13 & 56343.53 &      -  &     -  &  20.17 (0.02) &  19.87 (0.13) &     19.22 (0.03) &  4 \\ % LCOGT-Ste
 20/02/13 & 56343.62 &   19.86 (0.10)   &  20.75 (0.11) &  20.29 (0.07) &  19.99 (0.06) &     19.19 (0.05) & 3 \\ % NTT
 21/02/13 & 56344.53 &      -  &     -  &  20.17 (0.02) &  19.85 (0.05) &     19.10 (0.03) &  4 \\ %  LCOGT-Ste
 21/02/13 & 56344.57 &      -  &     -  &     -  &     -  &     19.14 (0.04) & 1 \\ 
 22/02/13 & 56345.53 &      -  &     -  &  20.11 (0.03) &  19.97 (0.05) &     19.20 (0.02) &  4 \\ %  LCOGT-Ste     
 23/02/13 & 56346.53 &      -  &     -  &  20.10 (0.08) &  20.01 (0.35) &           -      &  4 \\ %  LCOGT-Ste
 24/02/13 & 56347.53 &      -  &     -  &  19.96 (0.03) &  19.89 (0.13) &     19.24 (0.22) &  4 \\ %  LCOGT-Ste
 25/02/13 & 56348.53 &      -  &     -  &  20.00 (0.01) &  19.86 (0.08) &     19.23 (0.03) &  4 \\ %  LCOGT-Ste
 25/02/13 & 56348.55 &      -  &     -  &     -  &     -  &     19.17 (0.07) & 1 \\ 
 26/02/13 & 56349.53 &      -  &     -  &  20.09 (0.02) &  19.96 (0.02) &     19.23 (0.03) &  4 \\ % LCOGT-Ste
 28/02/13 & 56351.51 &      -  &     -  &  20.18 (0.15)  &     -  &     19.14 (0.07) & 1 \\ 
 01/03/13 & 56352.53 &      -  &     -  &  20.07 (0.03) &  20.00 (0.03) &     19.25 (0.03) &  4 \\ %  LCOGT-Ste
 01/03/13 & 56352.55 &      -  &     -  &     -  &     -  &     19.35 (0.08) & 1 \\
 02/03/13 & 56353.56 &      -  &     -  &     -  &     -  &     19.36 (0.07) & 1 \\ 
 03/03/13 & 56354.53 &      -  &     -  &     -  &     -  &     19.22 (0.07) & 1 \\ 
 03/03/13 & 56354.53 &   19.96 (0.04) &  20.78 (0.04) &  20.32 (0.03) &  20.02 (0.03) &     19.32 (0.02) & 3 \\ % NTT-photo
 05/03/13 & 56356.53 &      -  &     -  &  20.29 (0.01) &  20.09 (0.01) &     19.45 (0.11) & 4 \\ % LCOGT-Ste
 09/03/13 & 56360.53 &      -  &     -  &  20.23 (0.05) &  20.02 (0.01) &     19.24 (0.03) & 4 \\ % LCOGT-Ste
 11/03/13 & 56362.52 &      -  &     -  &     -  &     -  &     19.33 (0.06) & 1 \\
 12/03/13 & 56363.50 &      -  &     -  &     -  &     -  &     19.42 (0.13) & 1 \\
 14/03/13 & 56365.53 &      -  &     -  &  20.24 (0.02) &  19.97 (0.05) &     19.34 (0.03) & 5 \\ % LCOGT-Ste-1m0-09  
 17/03/13 & 56368.50 &      -  &     -  &     -  &     -  &     19.43 (0.08) & 1 \\
 17/03/13 & 56368.57 &   20.14 (0.07) &  20.88 (0.09) &  20.45 (0.09) &     -  &           -        & 3 \\ % NTT-photo    
 18/03/13 & 56369.54 &      -  &     -  &     -  &     -  &     19.36 (0.05) & 1 \\
 19/03/13 & 56370.50 &      -  &     -  &     -  &     -  &     19.36 (0.11) & 1 \\
 20/03/13 & 56372.49 &      -  &     -  &     -  &     -  &     19.38 (0.09) & 1 \\
 23/03/13 & 56374.51 &      -  &     -  &     -  &     -  &     19.62 (0.09) & 1 \\
 25/03/13 & 56376.51 &      -  &     -  &     -  &     -  &     19.64 (0.13) & 1 \\ \hline
\\
\caption{continued.} \\
\hline\hline
Date & JD & $U$ & $B$ & $V$ & $R$ & $I$ & Instrument \\
(dd/mm/yy) & (+2400000) & & & & & & \\ \hline
 03/04/13 & 56385.51 &   20.37 (0.04) &  21.08 (0.05) &  20.65 (0.04) & 20.36 (0.04) & 19.69 (0.03) &  3 \\
 03/04/13 & 56386.49 &      -  &     -  &     -  &     -  &     19.64 (0.09) & 1 \\ 
 06/04/13 & 56388.51 &   20.40 (0.10) &  21.12 (0.09) &  20.69 (0.06) & 20.42 (0.04) & 19.73 (0.05) &  3 \\ 
 14/04/13 & 56396.50 &      -  &  21.32 (0.10) &   20.81 (0.05) &    20.50 (0.05) &  19.80 (0.09) & 3 \\
 18/04/13 & 56401.49 &   20.74 (0.16) &    -   &    -   &     -   &      -   & 3 \\
 19/04/13 & 56402.48 &     -   &    -   &  20.98 (0.20) &     -   &      -   & 3 \\  
 01/08/13 & 56505.90 &      -  &     -  &     -  &     -  &     21.86 (0.55) & 1 \\
 10/08/13 & 56514.85 &      -  &     -  &     -  &     -  &    $>$ 22.85 & 1 \\
 17/08/13 & 56521.80 &      -  &     -  &     -  &     -  &    $>$ 23.04 & 1 \\
 20/08/13 & 56524.88 &      -  &     -  &     -  &     -  &    $>$ 22.71 & 1 \\ \hline
\end{longtable}
\begin{centering}
1 = 1.3m Warsaw University Telescope + OGLE-IV mosaic camera (Las Campanas Observatory);
2 = 2.2m MPI/ESO Telescope + GROND (ESO-La Silla Observatory);
3 = 3.58m New Technology Telescope + EFOSC2 (ESO-La Silla Observatory);
4 = LCOGT 1.0m-05 telescope + SBIG STX-16083 with Kodak KAF-16803 FI (Cerro Tololo Inter-American Observatory);
5 = LCOGT 1.0m-09 telescope + SBIG STX-16083 with Kodak KAF-16803 FI (Cerro Tololo Inter-American Observatory).
\end{centering}

\twocolumn

\begin{table*}
\caption{Near-infrared photometry of OGLE-2012-SN-006.}
\begin{center}
\begin{tabular}{cccccc} \hline\hline
Date & JD & $J$ & $H$ & $K$ & Instrument \\
(dd/mm/yy) & (+2400000) & & & & \\ \hline
31/10/12 & 56231.87 & 16.82  (0.06) & 15.77 (0.06) & 15.02 (0.08) & 1 \\ %grond
29/01/13 & 56321.57 & 17.62  (0.06) & 16.54 (0.05) & 15.50 (0.07) & 2 \\ % NTT
08/02/13 & 56331.61 & 17.81  (0.09) &   -          &   -          & 2 \\ % NTT
22/02/13 & 56345.73 & 17.92  (0.08) & 17.14 (0.18) & 15.96 (0.11) & 2 \\ % NTT
13/03/13 & 56364.51 & 18.18  (0.07) & 17.62 (0.10) & 16.23 (0.08) & 2 \\ % NTT 
12/04/13 & 56394.51 & 18.50  (0.10) & 17.70 (0.11) & 16.41 (0.09) & 2 \\ % NTT  
18/04/13 & 56400.53 & 18.51  (0.06) & 17.68 (0.07) & 16.43 (0.07) & 2 \\ % NTT   
\hline
\end{tabular}

1 = 2.2m MPI/ESO Telescope + GROND (ESO-La Silla Observatory);\\
2 = 3.58m New Technology Telescope + SOFI (ESO-La Silla Observatory).
\end{center}
\label{tabA2}
\end{table*}

\begin{table*}
\scriptsize
\caption{Magnitudes of the sequence stars in the field of OGLE-2012-SN-006. For the optical bands, errors in brackets are the
r.m.s. of the magnitudes obtained averaging estimates obtained in photometric nights. The NIR magnitudes and the associated errors are those
reported in the 2MASS catalogue \protect\citep{skru06}.}
\begin{center}
\begin{tabular}{ccccccccc} \hline\hline
Star & $U$ & $B$ & $V$ & $R$ & $I$ & $J$ & $H$ & $K$ \\ \hline
1  & 20.784 (0.035) & 19.282 (0.021) & 17.515 (0.012) & 16.195 (0.010) & 14.518 (0.020) & 12.919 (0.027) & 12.338 (0.022) & 11.983 (0.019) \\
2  & 20.446 (0.025) & 20.220 (0.059) & 19.417 (0.014) & 18.979 (0.012) & 18.523 (0.023) &   -            &   -            &   -            \\
3  & 18.712 (0.030) & 18.848 (0.011) & 18.129 (0.005) & 17.709 (0.007) & 17.298 (0.017) &   -            &   -            &   -            \\
4  & 17.637 (0.064) & 17.172 (0.172) & 16.289 (0.004) & 15.789 (0.008) & 15.358 (0.021) & 14.800 (0.047) & 14.313 (0.053) & 14.243 (0.088) \\
5  &  -             & 20.322 (0.020) & 18.525 (0.004) & 17.324 (0.011) & 15.682 (0.056) & 14.262 (0.032) & 13.634 (0.035) & 13.390 (0.038) \\
6  &  -             & 20.464 (0.045) & 18.827 (0.017) & 17.572 (0.040) & 16.112 (0.049) & 14.489 (0.048) & 13.830 (0.058) & 13.480 (0.125) \\
7  & 18.429 (0.023) & 17.127 (0.013) & 15.620 (0.004) & 14.699 (0.007) & 13.705 (0.050) & 12.650 (0.027) & 11.949 (0.023) & 11.762 (0.026) \\ 
8  & 20.746 (0.040) & 19.329 (0.021) & 17.779 (0.006) & 16.782 (0.009) & 15.630 (0.031) & 14.458 (0.029) & 13.791 (0.033) & 13.620 (0.048) \\
9  & 21.164 (0.023) & 19.936 (0.017) & 18.355 (0.008) & 17.281 (0.009) & 15.925 (0.022) & 14.668 (0.045) & 14.115 (0.048) & 13.729 (0.061) \\
10 & 21.419 (0.055) & 19.989 (0.017) & 18.411 (0.004) & 17.496 (0.009) & 16.534 (0.034) & 15.512 (0.067) & 14.916 (0.074) & 14.536 (0.108) \\
11 & 21.577 (0.072) & 20.554 (0.018) & 18.998 (0.006) & 18.089 (0.024) & 17.147 (0.026) & 15.997 (0.115) & 15.312 (0.116) & 15.075 (0.176) \\
12 &  -             & 21.546 (0.051) & 20.086 (0.015) & 19.159 (0.037) & 18.239 (0.025) &   -            &   -            &   -            \\
13 &  -             &  -             & 21.393 (0.046) & 20.286 (0.064) & 19.108 (0.110) &   -            &   -            &   -            \\
14 &   -            & 21.636 (0.020) & 20.241 (0.006) & 19.318 (0.021) & 18.530 (0.047) &   -            &   -            &   -            \\
15 &   -            &    -           & 21.983 (0.097) & 21.084 (0.048) & 20.258 (0.027) &   -            &   -            &   -            \\
16 & 16.899 (0.022) & 16.477 (0.007) & 15.621 (0.005) & 15.159 (0.008) & 14.723 (0.022) & 14.116 (0.045) & 13.744 (0.036) & 13.616 (0.048) \\
17 & 18.908 (0.028) & 17.739 (0.009) & 16.600 (0.004) & 15.926 (0.009) & 15.359 (0.022) & 14.588 (0.039) & 14.045 (0.039) & 13.823 (0.060) \\
18 & 20.246 (0.077) & 20.504 (0.017) & 19.852 (0.019) & 19.455 (0.022) & 19.070 (0.029) &   -            &   -            &   -            \\
19 &   -            &    -           & 21.135 (0.029) & 20.587 (0.030) & 20.076 (0.043) &   -            &   -            &   -            \\
20 & 18.925 (0.069) & 18.909 (0.017) & 18.493 (0.006) & 18.241 (0.007) & 18.022 (0.050) &   -            &   -            &   -            \\
21 & 17.614 (0.028) & 17.417 (0.016) & 16.590 (0.004) & 16.135 (0.011) & 15.716 (0.022) & 15.172 (0.058) & 14.934 (0.086) & 14.414 (0.093) \\
22 & 21.755 (0.166) & 20.450 (0.028) & 18.882 (0.012) & 17.770 (0.010) & 16.352 (0.056) & 15.026 (0.055) & 14.295 (0.052) & 14.173 (0.072) \\ 
23 & 15.584 (0.062) & 15.128 (0.013) & 14.215 (0.005) & 13.745 (0.009) & 13.351 (0.026) & 12.770 (0.029) & 12.371 (0.023) & 12.278 (0.026) \\
\hline
\end{tabular}
\end{center}
\label{tabA3}
\end{table*}

%\bsp

%\label{lastpage}


\begin{thebibliography}{}
\bibitem[Anupama et al.(2009)]{anu09} Anupama, G. C., Sahu, D. K., Gurugubelli, U. K., Prabhu, T. P., Tominaga, N., Tanaka, M., Nomoto, K. 2009, \mnras, 392, 894
\bibitem[Arnett(1982)]{arn82} Arnett W. D., 1982, \apj, 253, 785
\bibitem[Arnett(1996)]{arn96} Arnett, W. D. 1996, in {\it Supernovae and Nucleosynthesis: An Investigation of the History of Matter, from the Big Bang to the Present}, 
by D. Arnett. Princeton: Princeton University Press
\bibitem[Cappellaro et al.(1997)]{cap97} Cappellaro E., Mazzali P. A., Benetti S., Danziger I. J., Turatto M., della Valle M., Patat F. 1997, \aap, 328, 203
\bibitem[Chugai(2009)]{chu09} Chugai, N. N. 2009, \mnras, 400, 866
\bibitem[Clocchiatti \& Wheeler(1997)]{clo97} Clocchiatti A., Wheeler J. C. 1997, \apj, 491, 375
\bibitem[Di Carlo et al.(2008)]{dic08} Di Carlo, E. et al. 2008, \apj, 684, 471 
\bibitem[Falk \& Arnett(1977)]{falk77} Falk, S. W., Arnett, W. D. 1977, \apjs, 33, 151
\bibitem[Fisher(2000)]{fis00} Fisher, A. 2000, PhD Thesis, University of Oklahoma
\bibitem[Foley et al.(2007)]{fol07} Foley, R. J. et al. 2007, \apj, 657, L105 
\bibitem[Fransson et al.(2014)]{fra14} Fransson et al. 2014, \mnras, 797, 118
\bibitem[Fransson et al.(2005)]{fra05} Fransson et al. 2005, \apj, 622, 991
\bibitem[Gerardy et al.(2000)]{ger00} Gerardy, C. L., Fesen, R. A., H\"oflich, P., Wheeler, J. C. 2000, \aj, 119, 2968
\bibitem[Gorbikov et al.(2014)]{gor13} Gorbikov et al. 2014, \mnras, 443, 671
\bibitem[Greiner et al.(2008)]{gre08} Greiner et al. 2008, \pasp, 120, 405
\bibitem[Kozlowski et al.(2013)]{koz13} Koz{\l}owski, S. et al. 2013, Acta Astron., 63, 1
\bibitem[Landolt (1992)]{lan92} Landolt, A. U. 1992, \aj, 104, 340
\bibitem[Mattila et al.(2008)]{seppo08} Mattila, S. et al. 2008, \mnras, 389, 141 
\bibitem[Modjaz et al.(2009)]{mod09} Modjaz, M. et al. 2009, \apj, 702, 226                                                           
\bibitem[Moriya et al.(2011)]{mor11} Moriya, T. et al. 2011, \mnras, 415, 199
\bibitem[Moriya \& Tominaga(2012)]{mor12} Moriya, T. J., Tominaga, N. 2012, \apj, 747, 118
\bibitem[Nakano et al.(2006)]{nak06} Nakano, S., Itagaki, K., Puckett, T., Gorelli, R. 2006, CBET, 666, 1
\bibitem[Nozawa et al.(2008)]{noz08} Nozawa, T. et al. 2008, \apj, 684, 1343
\bibitem[Paczynski(1986)]{pac86} Paczynski, B. 1986, \apj, 304, 1
\bibitem[Paczynski(1991)]{pac91} Paczynski, B. 1991, \apj, 371, L63
\bibitem[Pastorello et al.(2007)]{pasto07} Pastorello, A. et al. 2007, \nat, 447, 829 
\bibitem[Pastorello et al.(2008a)]{pasto08b} Pastorello, A. et al. 2008a, \mnras, 389, 113 
\bibitem[Pastorello et al.(2008b)]{pasto08a} Pastorello, A. et al. 2008b, \mnras, 389, 131 
\bibitem[Pastorello et al.(2010)]{pasto10} Pastorello, A. et al. 2010, \apjl, 724, 16
\bibitem[Pastorello et al.(2015a)]{pasto13a} Pastorello, A. et al. 2015a, accepted (eprint arXiv: submit/1186154)
\bibitem[Pastorello et al.(2015b)]{pasto13b} Pastorello, A. et al. 2015b, accepted (eprint arXiv: submit/1186334) 
\bibitem[Prieto \& Morrell(2013)]{pri13}  Prieto, J. L., Morrell, N. 2012, Astron. Tel. 4734, 1
\bibitem[Quimby et al.(2007)]{qui07} Quimby, R. M., Aldering, G., Wheeler, J. C., H\"oflich, P., Akerlof, C. W., Rykoff, E. S. 2007, \apjl, 668, 99
\bibitem[Quimby et al.(2011)]{qui11} Quimby, R. M. et al. 2011, \nat, 474, 487
\bibitem[Sakon et al.(2009)]{sak09} Sakon, I. et al. 2009, \apj, 692, 546
\bibitem[Sanders et al.(2013)]{san13} Sanders, N. E. et al. 2013, \apj, 769, 39
\bibitem[Schlafly \& Finkbeiner (2011)]{sch11}  Schlafly, E. F., Finkbeiner D. P. 2011, \apj, 737, 103
\bibitem[Skrutskie et al.(2006)]{skru06} Skrutskie, M. F. et al. 2006, \aj, 131, 1163
\bibitem[Smartt et al.(2013)]{sma13} Smartt, S. J. et al. 2013, The Messenger, 154, 50
\bibitem[Smith et al.(2008)]{smi08} Smith, N., Foley, R. J., Filippenko, A. V. 2008, \apj, 680, 568
\bibitem[Smith et al.(2009)]{smi09} Smith, N. et al. 2009, \apj, 695, 1334
\bibitem[Smith et al.(2012)]{smi12} Smith, N. et al. 2012, \mnras, 426, 1905
\bibitem[Stoll et al.(2011)]{sto11} Stoll, R. et al. 2011, \apj, 730, 34
\bibitem[Stritzinger et al.(2012)]{str12} Stritzinger, M. et al. 2012, \apj, 756, 173
\bibitem[Taubenberger et al.(2013)]{tau13} Taubenberger, S. et al. 2013, \mnras, 432, 3117
\bibitem[Tominaga et al.(2008)]{tom08} Tominaga, N. et al. 2008, \apj, 687, 1208
\bibitem[Valenti et al.(2008)]{val08} Valenti S. et al. 2008, \mnras, 383, 1485
\bibitem[Udalski et al.(1997)]{uda97} Udalski, A., Kubiak M., Szyma\'nski, M. 1997,  Acta Astron., 47, 319
\bibitem[Wheeler et al.(2014)]{whee14} Wheeler, J. C, Johnson, V., Clocchiatti, A. 2014, \mnras, submitted (eprint arXiv:1411.5975)
\bibitem[Wozniak(2000)]{woz00} Wo\'zniak, P. R. 2000, Acta Astron., 50, 421
\bibitem[Wyrzykowski et al.(2012)]{wry12} Wyrzykowski, {\L}., Udalski, A., Koz{\l}owski, S. 2012, Astron. Tel. 4495, 1
\bibitem[Wyrzykowski et al.(2014)]{wyr14} Wyrzykowski, {\L}. et al. 2014, Acta Astronomica, submitted 
\bibitem[Yamaoka et al.(2006)]{yam06} Yamaoka, H., Nakano, S., Itagaki, K. 2006, CBET, 666, 2
\bibitem[Yaron \& Gal-Yam(2012)]{yar12} Yaron, O., Gal-Yam, A. 2012, \pasp, 124, 668
\bibitem[Zhang et al.(2012)]{zha12} Zhang, T. et al. 2012, \aj, 144, 131
\end{thebibliography}
\end{document}